\documentclass[11pt,a4paper]{article}

\usepackage{jheppub}
\usepackage{preamblejhep}
\usepackage{float}
\usepackage{epstopdf}

\usepackage{amsmath}
\usepackage{amsthm}
\theoremstyle{definition} 
\newtheorem{thm}{Theorem}

\theoremstyle{definition}

\newtheorem{defn}[thm]{Definition}
 
\usetikzlibrary{decorations.pathreplacing}

\begin{document}
\date{\today}
\title{Massless limit of massive self-interacting vector fields}

\author{Nabamita Banerjee,}
\author{Jitesh Singh}
\affiliation{
Department of Physics, Indian Institute of Science Education and Research Bhopal, Bhopal Bypass Road, Bhopal, 462066, India}

\emailAdd{nabamita@iiserb.ac.in}
\emailAdd{jitesh20@iiserb.ac.in}

\abstract{
We study massive self-interacting vector field theories with mass added ``by hand".
 We show that the massless limit of the quartic self-interacting vector field theory is not smooth.  Pathological behavior of the theory is not limited only at the classical level, even at the quantum level unitarity is violated at the two-loop. Using the Vainshtein mechanism, we show that it fails beyond the strong-coupling scale and hence a massless limit is not smooth even in quantum theory.
 }
\maketitle

\section{Introduction}
The generation of mass in quantum field theories can be approached through two distinct methods: the Higgs mechanism~\cite{Higgs1964} and the addition ``by hand''. The Higgs mechanism~\cite{Higgs1964}, which is a cornerstone of the Standard Model, ensures renormalizability by spontaneously breaking gauge symmetries~\cite{Lee1970,Boulware1970}. In contrast, adding mass ``by hand'' often leads to non-renormalizable theories and can violate the Ward identities~\cite{Ktorides}.
\\
\\
\textbf{Pathology at classical level:} The Proca theory~\cite{Proca1936}, which describes a massive vector field, provides a well-understood framework for analyzing the possibility of a non-zero vector or gauge boson mass. The application of self-interacting vector field is also widely used in gravity and cosmology~\cite{Esposito-Farese:2009wbc,DeFelice:2016yws,DeFelice:2016cri,Heisenberg:2017hwb,Kase:2017egk,Ramazanoglu:2017xbl,Annulli:2019fzq,Barton:2021wfj,Minamitsuji:2018kof, Herdeiro:2020jzx,Herdeiro:2021lwl}, inflationary dynamics  \cite{Dimopoulos:2006ms,Golovnev:2008cf,Maleknejad:2011jw,Maleknejad:2011sq,Adshead:2012kp,Emami:2016ldl,Garnica:2021fuu} and dark matter theories \cite{Holdom:1985ag,Graham:2015rva,Agrawal:2018vin,Goodsell:2009xc,Bastero-Gil:2018uel,Ema:2019yrd,Kolb:2020fwh,Ahmed:2020fhc,Caputo:2021eaa,Antypas:2022asj}. Problems in massive self-interacting vector fields where mass is added ``by hand" are not only limited to the non-renormalizability of the theory. Recently a few groups have utilized hyperbolic formulation and have shown that interaction beyond the Proca limit, are not well-posed \footnote{A well-posed problem is characterized by the existence of a solution, uniqueness of the solution, and continuous dependence on initial conditions.} due to the formation of singularities in 1+1 dimensions and the loss of hyperbolicity in higher dimensions~\cite{PhysRevLett.129.151103,PhysRevD.108.044022}. This loss of hyperbolicity implies that the classical equations of motion (EOM) change from hyperbolic (wave-like) to elliptical form, leading to ill-posedness~\cite{PhysRevD.108.044022}. We took the massless limit of these self-interacting theories and showed that they are also not smooth at the classical level. \\

\noindent The principle of continuity suggests that altering a physical theory should result in observables smoothly approaching their original values as the modification is reversed~\cite{SchrodingerBass1955}. However, massive self-interacting gauge theories with mass added ``by hand" appear to violate this principle. It is fascinating how a minor modification in the theory drastically affects its behavior.\\

\noindent \textit{In this paper one of our primarily goal is to understand the fate of the pathology under the massless limit}.\\ 

\noindent Quantum correction can avoid the classical singularity (kink) by smoothing it. The natural question is, \textit{does this pathological behavior persist at the quantum level?} To answer this question we check unitarity of the theory in the massless limit at the scale where the equations of motion suffer from hyperbolicity loss.
\\
\\
\textbf{Pathology at Quantum level:} Conventional approaches indicate that these theories do not have a smooth massless limit, with perturbative series becoming singular in mass, suggesting a violation of unitarity or failure of the regular perturbation theory~\cite{Hell2022}. This issue is not unique to vector fields; linearized massive gravity also suffers from a similar problem known as the van Dam-Veltman-Zakharov (vDVZ) discontinuity~\cite{vanDam1970}. This discontinuity arises because the longitudinal mode in massive gravity fails to decouple from the other degrees of freedom as the mass approaches zero. Both General Relativity and massive gravity are nonlinear theories that are typically solved using perturbation theory. However, there exists a scale known as the Vainshtein radius~\cite{Vainshtein1972gravity}, at which nonlinear terms become significant, causing perturbation theory to break down for longitudinal modes~\cite{Vainshtein1972,Vainshtein1972gravity}. This breakdown leads to a strong coupling regime in which the longitudinal mode decouples, restoring General Relativity's predictions in the massless limit~\cite{Vainshtein1972gravity}. Recent work attempted to demonstrate a smooth massless limit for massive self-interacting theories using the Vainshtein mechanism~\cite{Hell2022}. However, our findings using the hyperbolic formulation suggest that the massless limit is not smooth at the classical level, in presence of quartic self coupling. We revisited the non-covariant formulation and compared it with the hyperbolic formulation, which helps in determining the true behavior of the massless limit at the quantum level.\\

\noindent
The plan of the paper is as follows: in section \ref{sec2} we have introduced the non-linear Proca theory (NPL) with quartic self-interaction and discuss the hyperbolic formulation of its equation of motion. We show that the hyperbolicity of the equation is lost at a finite value of the field amplitude. Next in section \ref{sec3} we discuss the perturbative classical study of the theory. In section \ref{sec4}, we present a non-covariant formulation of the theory, that is useful for the discussions in the later sections. Section \ref{sec5} discusses the unitarity violation in the corresponding quantum theory and in sections \ref{sec6} and \ref{sec7} we describe the unitarity violation as an artifact of the breakdown of the corresponding perturbative approach. We explained the energy scales where the theory is well approximated. The implications of the massless limit has been discussed in all sections. The paper ends with a discussion in section \ref{sec8}. Necessary technicalities of the cubic and quartic NPL are presented in the first four appendices, whereas the last appendix contains results for massive Yang-Mills theory. Unlike massive abelian vector field theories, massive Yang-Mills, a theory of non-abelian vector field in which two types of intrinsic self-interactions are present: quartic and cubic derivative self interaction, has well posed classical EOM due to the internal symmetry. This has been recently shown in Ref.~\cite{jose2025internalsymmetryrescuewellposed} \footnote{This can be immediately verified by setting the additional interaction parameter to zero.}.\\
\section{Hyperbolic Formulation of Nonlinear Proca Theory}\label{sec2}
The signature of the space-time metric is $(+,-,\dots,-)$.\\
\\
In this section, we review the hyperbolic formulation for vector fields. We start with the Proca theory, which describes a massive vector field, and then proceed to self-interacting theories. We utilize the hyperbolic formulation to check the hyperbolicity of the equation of motion. The Lagrangian and the action for the Proca theory in a fixed space time with metric $g_{\mu\nu}$ is given by
\begin{align}\label{eq:procaaction}
    {\cal{L}}_{PT} = -\frac{1}{4} F_{\mu\nu}F^{\mu\nu} +  \frac{\mu^2}{2} X^2 \ \quad\text{and,}\quad {\cal{S}}_{PT} =\int d^4x \sqrt{-g} \left(-\frac{1}{4} F_{\mu\nu}F^{\mu\nu} +  \frac{\mu^2}{2} X^2\right)
\end{align}
where $X^2 =X_\mu X^\mu$ and $F_{\mu\nu} = \nabla_\mu X_\nu - \nabla_\nu X_\mu$ for the real vector field $X_\mu$. The corresponding field equation after utilizing the on-shell Lorenz condition (\(\nabla_\mu X^\mu=0\)) is  
\begin{align}\label{eompt}
     g_{\mu\rho}  \nabla^\mu \nabla^\rho X_\nu -  R_{\mu \nu} X^\mu + \mu^2 X_\nu = 0 ; \quad \Box=g^{\mu\nu}\nabla_{\nu} \nabla_{\mu},
\end{align}
where $R_{\mu \nu}$ is Ricci tensor. For Lorentzian space-time metric $g_{\mu \nu}$, the equation of motion (\ref{eompt}) is hyperbolic (wave-like). In case of self-interacting theories, this is not true. Let us start with a massive vector field with quartic self-interaction     
\begin{align}\label{eq:action}
    {\cal{L}}_{NPL} = -\frac{1}{4} F_{\mu\nu}F^{\mu\nu} + \left( \frac{\mu^2}{2} X^2 + \frac{\lambda}{4} \left( X^2\right)^2 \right) \ ,
\end{align}
where $X^2 =X_\mu X^\mu$ and  $F_{\mu\nu} = \nabla_\mu X_\nu - \nabla_\nu X_\mu$for the real vector field $X_\mu$. We call this Nonlinear Proca theory (NPL).
The corresponding field equation after utilizing the on-shell Lorenz condition (\(\nabla_\mu (zX^\mu)=0\)) is  ~\cite{PhysRevLett.129.151103}
\begin{align}
    \label{eq:eomnpt}
    \bar{g}_{\mu\rho} \nabla^\mu \nabla^\rho X_\nu& + 2z'z^{-1} \Theta_\nu  - R_{\mu \nu} X^\mu - z  X_\nu + \cdots = 0 \\
    \text{where, }\bar{g}_{\mu\rho} &= g_{\mu\rho} + 2z'z^{-1} X_\mu X_\rho , \Theta_{\nu}= X^\mu X^\rho \nabla_\mu F_{\nu\rho}\nonumber\\
    z&=-\left(\mu^2+\lambda X^2\right) \text{ and, } z'=\frac{dz}{dX^2}.
\end{align}
Where ellipses represent the product of single derivative terms. Since the EOM is a second-order differential equation, it can be classified based on the signature of the principal part of the differential equation.
For the $1+1$ dimension, $\Theta_\nu = 0$ ~\cite{PhysRevLett.129.151103}, the EOM is controlled and classified solely by the effective metric $\bar{g}_{\mu\rho}$ and is given as,
\begin{align}\label{eq:gbar_eom}
    \bar{g}_{\alpha\beta}\nabla^\alpha \nabla^\beta X^\mu + \dots 
    = {\cal M}^{\mu}{}_{\alpha} X^\alpha
\end{align}
where ${\cal M}_{\mu v}=-\mu^2 \delta_{\mu v}+R_{\mu v}$ is the mass matrix. 
Let us now state two important definitions and a theorem that are necessary for our study. 
\defn{
    A Cauchy problem is called well-posed if there exist constants $K$ and $\alpha$ such that
\begin{align}\label{eqnwell}||exp(P(ik)t)||\leqslant Kexp(\alpha t) \quad \forall t>0 \ .
\end{align}
Where $P(ik)$ is the principal symbol of a PDE~\cite{Sarbach_2012}.
} The significance of this definition is that it is based on the property that for each fixed time $t>0$ the norm $\left|e^{P(i k) t}\right|$ of the propagator is bounded by the constant $f(t):=K e^{\alpha t}$, which is independent of the wave vector $k$~\cite{Sarbach_2012}.
\thm{\label{theo1}
Consider a Cauchy problem of a second-order hyperbolic equation:
\begin{equation}    \begin{cases}\left(\partial_t^2-a(t)^2 \Delta\right) u(t, x)=0, & (t, x) \in[0, T] \times \mathbb{R}^n \\ u(0, x)=u_0(x), u_t(T, x)=u_1(x), & x \in \mathbb{R}^n
\end{cases}
\label{cauchy1}
\end{equation}
where $\Delta=\sum_{j=1}^n \partial_{x_j}^2, a(t) \geq 0$ and $T$ is a small positive number. If $a(t)$ has a singularity, which means non-Lipschitz continuity\footnote{A real-valued function \(f: \mathbb{R} \to \mathbb{R}\) is called Lipschitz continuous if there exists a positive real constant $K$ such that, for all real $x_1$ and $x_2$, $\left| f(x_1) - f(x_2) \right| \leq K \left| x_1 - x_2 \right|$.} or having a zero, then $L^2$ well-posedness does not hold and it leads to loss of regularity~\cite{2003MMAS...26..783H}.
}\\

\noindent $\bar{g}_{\mu\nu}$ is a disformal metric, which in $1+1$ dimensions  locally can be written as ~\cite{Bittencourt2018,Bittencourt_2015}  
\begin{align}
&\bar{g}_{\mu\nu} = \operatorname{diag} ( \frac{z_3}{z}, -1  )\\
    &\text{where, } z_3=-\left(\mu^2+3\lambda X^2\right)\nonumber
\end{align}

\noindent Using the above theorem to the EOM for NPL in $1+1$ dimensions, we see that it exhibit a regularity loss. Here, $\frac{z}{z_3}$ acts as $a(t)$, which has a singularity at  $z_3=0$ and is zero at $z=0$. Consequently, equation of motion (\ref{eq:gbar_eom}) is not well-posed in $1+1$ dimensions.
\defn{
    (Loss of hyperbolicity)
    \\Starting with a small amplitude $\lambda X^2 << 1$, the effective metric is initially Lorentzian but alters its signature at a finite value of the field amplitude~\cite{PhysRevD.108.044022}. Thus, the field equations go from hyperbolic to elliptic. This is called dynamical loss of hyperbolicity~\cite{PhysRevD.108.044022}.
} The principal symbol $\mathcal{P}(k)$, which characterizes the dynamics is ~\cite{PhysRevLett.129.151103}
\begin{align}
    \mathcal{P}(k)^\beta{}_\alpha &= g_{\mu\nu}k^\mu k^\nu \delta^\beta_\alpha+2z'z^{-1}(X^\mu k_\mu)X^\beta k_\alpha. 
\end{align}
Determinant of the principal symbol is 
\begin{align}\label{eq:k-w}
    \det \mathcal{P}(k)&=\left( g_{\mu\nu}k^{\mu}k^{\nu}\right)^d\left(\bar{g}_{\alpha\beta}k^\alpha k^\beta\right)\ .
\end{align}
\noindent In $d+1$ dimensions, the principal part of differential Eq. (\ref{eq:eomnpt}) has two terms, one with effective metric $\bar{g}_{\mu\rho}$ and the other is $\Theta_\nu $. Hence, two different modes exist in higher dimensions. The mode controlled by the effective metric is 
\begin{align}\label{wk}
    \bar{g}_{\mu\nu} k^{\mu} k^{\nu}=0.
\end{align}
$\bar{g}_{\mu\nu}$ is a disformal metric, which locally can be written as ~\cite{Bittencourt2018,Bittencourt_2015}  
\[
\bar{g}_{\mu\nu} = \operatorname{diag} ( \frac{z_3}{z}, \underbrace{-1, -1, \dots, -1}_{d \text{ times}} ).
\]
As evident $\bar{g}_{\mu\nu}$ changes its signature for the finite value of field amplitude. This means that the frequency $\omega$ of the dispersion relation (\ref{wk}) becomes imaginary when the effective metric changes its signature from hyperbolic to elliptic~\cite{PhysRevD.108.044022}. In the mode expansion of vector field $e^{-i \omega t}$ becomes $e^{|\omega| t}$. Consequently, these modes increase exponentially rather than oscillating. Massive Yang-Mills does not suffer from the loss of hyperbolicity issue as described in Appendix \ref{appendix5}.
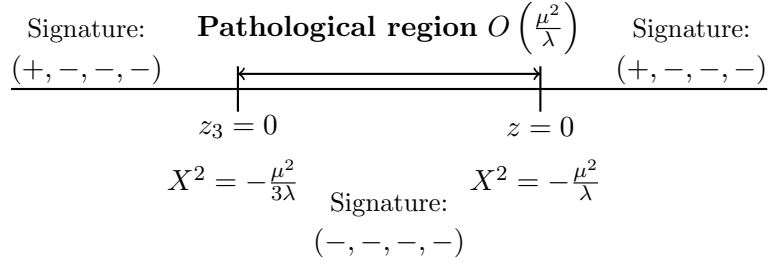
\begin{figure}[H]
    \centering
    \begin{tikzpicture}
        \draw[thick] (-5,0) -- (5,0); 

        \draw[thick] (-2,0.3) -- (-2,-0.3);
        \node at (-2,-0.5) {$z_3=0$};

        \draw[thick] (2,0.3) -- (2,-0.3);
        \node at (2,-0.5) {$z=0$};

        \draw[thick, <->] (-2,0.2) -- (2,0.2);
        \node[align=center] at (0,0.8) {\textbf{Pathological region} $O\left(\frac{\mu^2}{\lambda}\right)$};

        \node[align=left] at (-2,-1.2) {
            $X^2 = -\frac{\mu^2}{3\lambda}$
        };

        \node[align=left] at (2,-1.2) {
            $X^2 = -\frac{\mu^2}{\lambda}$
        };

        \node[align=center] at (-4,0.5) {\small Signature: \\ $(+,-,-,-)$};
        \node[align=center] at (4,0.5) {\small Signature: \\ $(+,-,-,-)$};
        \node[align=center] at (0,-1.8) {\small Signature: \\ $(-,-,-,-)$};
    \end{tikzpicture}
    \caption{Loss and restore of hyperbolicity for $\bar{g}_{\mu\rho} = g_{\mu\rho} + 2z'z^{-1} X_\mu X_\rho$ in $3+1$ dimensions.}
    \label{fig:pathological-region}
\end{figure}
\noindent In conclusion the massive self-interacting vector field theory is ill-posed  in any dimensions (i.e, $1+1$ or higher dimensions)~\cite{PhysRevLett.129.151103}. The $1+1$ dimensions case was chosen primarily to illustrate these phenomena more explicitly. In $1+1$ dimensions all modes are controlled by the effective metric which is a vector disformal vector metric ($\bar g_{\mu\nu}$) and it is degenerate. While in higher dimensions there are two different modes, one is controlled by the space time metric and another one is controlled by the effective metric ($\bar g_{\mu\nu}$) and it is the potential root cause of the loss of hyperbolicity~\cite{Hell2022}.   
\subsection{Without mass term}
Let us begin with the self-interacting term $\lambda X^4$, omitting the mass-term $ \mu^2 X^2$  
\begin{align}\label{action without mass term}
    {\cal{L}} =-\frac{1}{4} F_{\mu \nu} F^{\mu \nu}+\frac{\lambda}{4}\left(X^2\right)^2 .
\end{align}
The Equation of motion for Eq. (\ref{action without mass term}) after utilizing the generalized Lorenz condition as derived in Appendix \ref{appendix1} is
\begin{align}\label{eq:eeomm0}
    &\bar{g}_{\mu\rho} \nabla^\mu \nabla^\rho X_\nu - 2\lambda z_0'^{-1} \Theta_\nu  - R_{\mu \nu} X^\mu - z_0' X_\nu + \cdots = 0 \nonumber\\
    &\text{where, }\bar{g}_{\mu\rho} = g_{\mu\rho} - 2\lambda z_0'^{-1} X_\mu X_\rho \text{, } z_0'= -\lambda X^2 \text{ and }
    \nonumber\\
    & \qquad \quad \Theta_{\nu}= X^\mu X^\rho \nabla_\mu F_{\nu\rho}
\end{align}
where ellipses represent the product of single derivative terms.
Recall in $1+1$ dimensions $X^\mu X^\rho \nabla_\mu F_{\nu \rho}=0$. So, the only term that contributes to the principle part in 1+1 dimensions is controlled by $\bar{g}_{\mu\nu}$. Let's compute determinant of $\bar{g}_{\mu \rho}=g_{\mu\rho} - 2\lambda z_0'^{-1} X_\mu X_\rho$
$$
\begin{aligned}
\operatorname{det}\left(\bar{g}_{\mu \rho}\right) & =\operatorname{det}\left(g_{\mu \rho}\right)\left(1+X_\mu \frac{-2 \lambda}{z_0'} g^{\mu \rho} X_\rho\right) \\
\bar{g} & =g\left(1+\frac{2 \lambda X^2}{ \lambda X^2}\right)=3 g .
\end{aligned}
$$
Since $\operatorname{det}(\bar{g}_{\mu \rho})$ is non zero, there is no pathology. $\bar{g}_{\mu\nu}$ locally can be written as ~\cite{Bittencourt2018}
\begin{equation}
\bar{g}_{\mu\nu} = \operatorname{diag} ( 3, \underbrace{-1, -1, \dots, -1}_{d \text{ times}})
\end{equation}
in $d+1$ dimensions, also locally the $\bar{g}_{\mu\nu}$ is field independent.
Hence, there is no breakdown or loss of hyperbolicity for the vector field model without a mass term. Taking $\mu^2\to0$ to the massive case gives the massless case except for the configuration $z_3=0$ or $z=0$; hence, the calculation is consistent\footnote{We are working with $\bar{g}_{\mu\rho} =  g_{\mu\rho} + 2z'z^{-1} X_\mu X_\rho$ and have shown that there is no pathology for the massless case. But if we work with the scaled effective metric, the pathology remains. This is another reason that the two effective metrics are not equivalent.}. Thus we understand that the mass term is intrinsic for the appearance of the singularity. The purely massless self-interacting vector theory is non-singular, whereas the one with a mass term shows singularity even in the  $\mu \rightarrow 0 $ limit as we clearly show below. \\  

\subsection{Massless limit}
The effective metric that governs the dynamics of NPL is, 
\begin{align}
\bar{g}_{\mu\nu} &=
\operatorname{diag} ( \frac{z_3}{z}, \underbrace{-1, -1, \dots, -1}_{d \text{ times}} )\nonumber\\
&=
\operatorname{diag} ( \frac{\mu^2 + 3\lambda X^2}{\mu^2 + \lambda X^2}, \underbrace{-1, -1, \dots, -1}_{d \text{ times}} )\nonumber\\
 &\xrightarrow{\mu^2 \to 0} 
\operatorname{diag} ( 3, \underbrace{-1, -1, \dots, -1}_{d \text{ times}}).
\end{align}
The effective metric in this case is locally field amplitude independent, which guarantees well-posedness of the classical solution. Further from the local analysis of effective metrics, we understood the mathematical reason behind the disappearance of the pathology. Now, let's try to understand it by looking at another picture. For the massive vector field, the pathology emerges for values of the vector field of the order of $\mu/\sqrt{\lambda}$, and as we take the limit $\mu \to 0$, the region of field space between $z_3 = 0$ and $z = 0$ shrinks to $O(\mu^2)$ in the massless limit. However, we still do not get rid of the pathological configurations $z_3 = 0$ ($X_\mu X^\mu = \frac{-\mu^2}{3\lambda}$) and $z = 0$ ($X_\mu X^\mu = \frac{-\mu^2}{\lambda}$), which in the massless limit correspond to $z_3 \sim z \sim O(\mu^2)$. Hence, these configurations lead to regularity loss in the vector field solutions even for arbitrarily small mass. In the next section, we will discuss how these configurations affect the analysis. 
\begin{figure*}[t]
\centering
\begin{tikzpicture}[scale=0.8, transform shape]
\draw[thick] (-5,0) -- (5,0); 
\draw[thick] (-2,0.3) -- (-2,-0.3);
\node at (-2,-0.5) {$z_3=0$};
        \draw[thick] (2,0.3) -- (2,-0.3);
        \node at (2,-0.5) {$z=0$};
        \draw[thick, <->] (-2,0.2) -- (2,0.2);
        \node[align=center] at (0,2) {\textbf{Pathological region} \\ $O\left(\frac{\mu^2}{\lambda}\right)$};
        \node[align=left] at (-2,-1.2) {$X^2 = -\frac{\mu^2}{3\lambda}$};
        \node[align=left] at (2,-1.2) {$X^2 = -\frac{\mu^2}{\lambda}$};
        \node[align=center] at (-4,0.5) {\small Signature: \\ $(+,-,-,-)$};
        \node[align=center] at (4,0.5) {\small Signature: \\ $(+,-,-,-)$};
        \node[align=center] at (0,-1.8) {\small Signature: \\ $(-,-,-,-)$};

        \draw[thick, ->] (6,0) -- (8,0) node[midway, above] {\textbf{Massless limit}};

        \draw[thick] (9,0) -- (15,0); 
        \draw[thick] (12,0.3) -- (12,-0.3);
        \node at (11.3,-0.5) {$z_3=0$};
        \draw[thick] (12.1,0.3) -- (12.1,-0.3);
        \node at (12.7,-0.5) { $z=0$};
        \draw[thick, <->] (12.0,0.2) -- (12.1,0.2);
        \node[align=center] at (12,2) {\textbf{Pathological region} \\ $O\left(\frac{\mu^2}{\lambda}\right) \to 0$};
        \node[align=left] at (12,-1.2) {$X^2 \approx -\frac{\mu^2}{\lambda} \approx -\frac{\mu^2}{3\lambda}$};
        
        \node[align=center] at (10,0.5) {\small Signature: \\ $(+,-,-,-)$};
        \node[align=center] at (14,0.5) {\small Signature: \\ $(+,-,-,-)$};
    \end{tikzpicture}
    \caption{Transition to the massless limit: The pathological region shrinks as $O\left(\frac{\mu^2}{\lambda}\right) \to 0$ but we still have the pathological configurations left over $z_3\sim z\sim O(\mu^2)$.}
    \label{fig:pathological-region-transition}
\end{figure*}
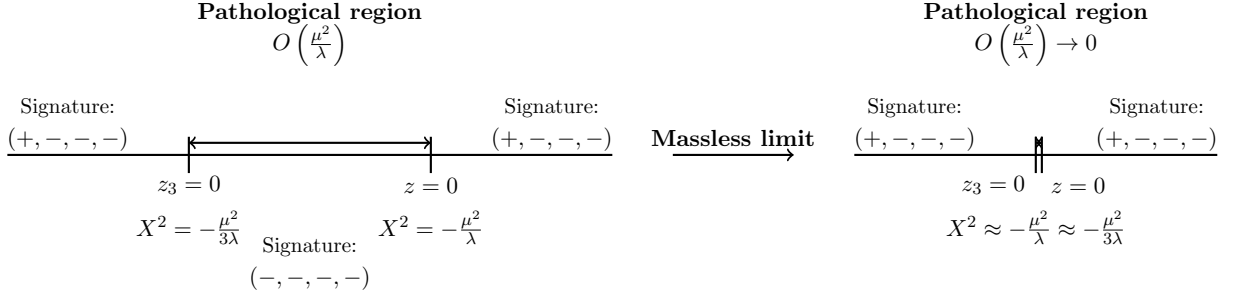
\section{Perturbative classical study of Nonlinear Proca Theory}\label{sec3}
The above analysis shows the pathology while treating the interaction term exactly. We may ask if the same is also evident in a perturbative analysis\footnote{We thank Akshay Kumar for useful discussions on this point.}. We address the same in this section. The equation of motion for (\ref{eq:action}) after utilizing the on-shell Lorenz condition reduces to 
\begin{align} \label{exacteom}
\bar{g}_{\mu \rho} \nabla^\mu \nabla^\rho X_v &+2 z' z^{-1} \theta_v-R_{\mu v} X^\mu-z X_v+\tilde{g}^{\mu \rho} \nabla_v X_\rho \nabla_\mu \ln z \nonumber\\
&+2 z' z^{-1} X^\mu \nabla_v X_\rho \nabla_\mu X^\rho=0.
\end{align}
Where,
$$
\begin{aligned}
\bar{g}_{\mu \rho} & =g_{\mu \rho}+2 z' z^{-1} X_\mu X_\rho, \\
\tilde{g}_{\mu \rho} & =g^{\mu \rho}-2 z' z^{-1} X^\rho X^\mu, \\
\theta_v & =X^\mu X^\rho \nabla_\mu F_{v \rho}.
\end{aligned}
$$
The perturbative expansion of the vector field in ``Regular perturbation theory" is
\begin{align}\label{pertubativeeom}
&X_\mu =X_\mu^0+\lambda \delta X_\mu+O\left(\lambda^2\right). 
\end{align}
Equation of motion at $O\left(\lambda^{0}\right)$ and $O\left(\lambda^{1}\right)$ in $1+1$ dimensions becomes:
$$
\begin{aligned}
& O\left(\lambda^0\right): \square X_v^0-M_{\mu v} X^{0 \mu}=0 \text { and } \\
& O\left(\lambda^{1}\right): \square \delta X_v-M_{\mu v} \delta X^\mu+J_v=0,
\end{aligned}
$$
where, $\square=g_{\mu \rho} \nabla^\mu \nabla^\rho$, $M_{\mu v}=-\mu^2 \delta_{\mu v}+R_{\mu v}$ and $J_v=\mu^{-2} g^{\mu \rho} \nabla_v X_\rho^0 \nabla_\mu \operatorname{ln}(-X^0)^{2}+2 \mu^{-2} X_\mu^0 X_\rho^0 \nabla^\mu \nabla^\rho X_v^0+ 2 \mu^{-2} X^{0 \mu} \nabla_v X_\rho^0 \nabla_\mu X^{0 \rho}+(X^0)^{2} X_v^0$.
\subsection{Local analysis}
To simplify the perturbative equation of motion we use the local analysis as described in Appendix \ref{appendix2}. For a global time like vector field, which is the solution of Proca theory (order zero equation), order by order solution is given by\\
Solution for $O\left(\lambda^0\right): \tilde{X}^0=A e^{-\mu x}+B e^{\mu x}$.\\
Solution for $O\left(\lambda^{1}\right): \delta X_{0 {\text{homo}}} +\delta X_{0 \text{particular}}$
$$
\begin{aligned}
= A^{\prime} e^{i(\omega t-k x)}&+B^{\prime} e^{-i(\omega t+k x)}+\frac{A^3}{8 \mu^2} e^{-3 \mu x}+\frac{B^3}{8 \mu^2} e^{3 \mu x}\\
-&\frac{3 A^2 B}{2 \mu} x e^{-\mu x}+\frac{3 A B^2}{2 \mu} x e^{\mu x}.
\end{aligned}
$$
Total solution is $X_\mu=X_\mu^0+\lambda \delta X_\mu+O\left(\lambda^2\right)$.\\
\\
Similarly, for a global space like vector field, the order-by-order solution is given by\\
Solution of $O\left(\lambda^0\right): \tilde{X}^{1}=A e^{i \mu t}+B e^{-i \mu t}$.\\
Solution of $O\left(\lambda^{1}\right): \delta X_1=\delta X_{1 \text{ homo}}+\delta X_{1 \text { particular }}$
$$
\begin{aligned}
=A^{\prime} e^{i(\omega t-k x)}&+B^{\prime} e^{-i(k x+\omega t)} +\frac{A^3}{8 \mu^2} e^{3 i \mu t}
-\frac{3 A^2 B}{2 i \mu} t e^{i \mu t}\\
&+\frac{3 B^2 A}{2 i \mu} t e^{-i \mu t} +\frac{B^3}{8 \mu^2} e^{-3 i \mu t}.
\end{aligned}
$$
The secular term: $-\frac{3 A^2 B}{2 i \mu} t e^{i \mu t}$ and $\frac{3 B^2 A}{2 i \mu} t e^{-i \mu t}$. \\
The total solution up to $O\left(\lambda^2\right)$ is  
$$
X^\mu=X^{0 \mu}+\lambda \delta X^\mu+O\left(\lambda^2\right).
$$
The secular term is proportional to $t$ and leads to an unbounded solution after timescales of $t \sim O\left(\frac{1}{\lambda}\right)$. These secular terms are artifacts of the Regular perturbation theory (RPT) and do not represent the true behavior of the solution.
\subsection{Asymptotic perturbation theory}
We use the Poincaré-Lindstedt method to remove the secular term from the solution. Let's define a stretched time variable $\tau$ as
$$
\begin{aligned}
\tau & =\omega(\lambda) t. \\
\end{aligned}
$$
Asymptotic expansion of the form
$$
\begin{aligned}
& X^\mu=X^{0\mu}+\lambda \delta X^\mu+O\left(\lambda^2\right) \\
& \omega(\lambda)=1+\lambda \omega_1+ O\left(\lambda^2\right).
\end{aligned}
$$
To remove the secular term  $\frac{3 A B}{2 \mu^2}+ \omega_1$ should be zero and for that $\omega_1=-\frac{3 A B}{2 \mu^2}$ as shown in Appendix \ref{appendix2}. Hence, the asymptotic expansion of the frequency becomes 
\begin{align}
    \omega=1-\frac{3 A B}{2 \mu^2} \lambda+O\left(\lambda^2\right).
\end{align}
The solution of NPL after using the asymptotic perturbation theory is regular.\\
\\
The Poincaré–Lindstedt method is used when regular perturbation theory fails due to the appearance of secular terms. For example, in the standard setting, regular perturbation theory breaks down when the nonlinear interaction terms become comparable in magnitude to the linear terms.

\subsection{Failure of asymptotic perturbation theory}
\textit{Why does the asymptotic perturbation theory give a regular solution while the exact solution suffers from the singularity issue?}\\
To understand the same, let us recall, the EOM for NPL is given by Eq. (\ref{exacteom})
$$
\begin{aligned}
&\bar{g}_{\mu \rho} \nabla^\mu \nabla^\rho X_v +2 z' z^{-1} \theta_v +2 z' z^{-1} X^\mu \nabla_v X_\rho \nabla_\mu X^\rho\\&+\tilde{g}^{\mu \rho} \nabla_v X_\rho \nabla_\mu \ln z-R_{\mu v} X^\mu-z X_v=0.
\end{aligned}
$$
In $1+1$ dimension, $\theta_v=0$ and the perturbative EOM using the perturbative expansion of the vector field in RPT is given by (\ref{pertubativeeom})
$$
\begin{aligned}
O\left(\lambda^0\right): & g_{\mu \rho} \nabla^\mu \nabla^\rho X_v^0+\mu^2 X_v^0-R_{\mu v} X^{0 \mu}=0 \text{ and, }\\
O\left(\lambda^1\right): & g_{\mu \rho} \nabla^\mu \nabla^\rho \delta X_v+\frac{2}{\mu^2} X^{0 \mu} \nabla_v X_\rho^0 \nabla_\mu X^{0 \rho}+\mu^2 \delta X_v\\
&+\frac{2}{\mu^2} X_\mu^0 X_\rho^0  \nabla^\mu \nabla^\rho X_v^0+\frac{g^{\mu \rho}}{\mu^2} \nabla_v X_\rho^0 \nabla_\mu \operatorname{ln}(-X^0)^2  \\
&-R_{\mu v} \delta X^\mu+{(X^0)}^2 X_v^0=0. 
\end{aligned}
$$
In the above zero-order perturbative equation, we presumed that the second term from the effective metric ``$2 \lambda z^{-1} X_\mu X_\rho$" never contributes, but this is not true for all field configurations; for example, $X^2 \sim O\left(\frac{\mu^2}{\lambda}\right)$ it contributes and changes the nature of the EOM. Thus, mathematically, while we consider all possible field configurations, the perturbation technique does not work. On the other hand, if we restrict to special field configurations, it holds. \\  
\\
Hence, this breakdown cannot be resolved even after including the non-linear terms from the interaction term (which is the working principle behind the Vainshtein mechanism). In our case, the situation is more severe. The principal part of the kinetic term vanishes at $z_3 = 0$ and $z = 0$, leading to the formation of kinks (i.e., loss of derivatives). This indicates a breakdown of the hyperbolic structure of the equations. Consequently, one cannot guarantee that the kinetic term remains suppressed beyond the Vainshtein radius\footnote{Vainshtein mechanism is used to claim the unitarity of the theory for the regime where the energy scale is much grater than the mass~\cite{Hell2022},as will be discussed in \ref{sec6}}, even if nonlinear effects are taken into account.\\
\\
The purpose of this section is to show that if the theory  exhibits a pathology (not well posed), then perturbative analysis fails. In the case, where we treat the quartic coupling perturbatively, the failure is quite trivial, as the Perturbative treatment breaks when the field take values of $O\left(\frac{1}{\sqrt{\lambda}}\right)$, which is quite large.\\
\\
Similarly, the massless theory is well-posed, but the massive theory, even with arbitrarily small mass, is not. Hence, if we perform a perturbative analysis with the mass ($\mu^2$) as the perturbative parameter, we again encounter a breakdown of the perturbative expansion due to the hyperbolicity loss. However, the key difference in this compared to the quartic coupling case is that hyperbolicity loss occurs at a finite, small value of the field amplitude: $|X| \sim O(\mu)$. This makes the situation non-trivial and can be understood from the previous perturbative analysis.
\section{A non-covariant formulation}\label{sec4}
This section briefly review the non-covariant formulation, is originally
discussed in ~\cite{Hell2022,Hell:2021YM}. For convenience, we follow the notation based on Ref.~\cite{Hell2022} in the upcoming sections. In this section, we express the free theories of Proca fields in their propagating degree of freedom~\cite{Demozzi2009,Hell2022,Hell_2022,Hell:2021YM} for flat space-time. 
We use commas to denote the derivative with respect to the corresponding coordinate $x^{\mu}$. The action for the Proca theory is given by
\begin{equation}
\label{part2-1}
S=\int d^4x\left(-\frac{1}{4}F_{\mu\nu}F^{\mu\nu}+\frac{\mu^2}{2}X_{\mu}X^{\mu}\right),
\end{equation}
where
\begin{equation}
\label{part2-2}
F_{\mu\nu}=X_{\nu,\mu}-X_{\mu,\nu}
\end{equation}
is the field strength tensor for the vector field $X_{\mu}$. We break down the spatial component of the vector field into longitudinal and transverse parts as
\begin{equation}
\label{part2-3}
X_i=X_i^T+\chi_{,i},\qquad\text{with}\qquad X_{i,i}^T=0.
\end{equation}
The transverse modes are denoted by \(X_i^T\), and the longitudinal mode by \(\chi\).$X_0$ is the non-propagating component in Proca theory. Taking the variation of the Proca action with respect to it yields a constraint
\begin{equation}
\label{part2-5}
(-\Delta+\mu^2)X^0=-\Delta\dot{\chi}.
\end{equation}
The action in terms of the propagating degrees of freedom, after applying the constraint, is
\begin{equation}
\label{part2-9}
S=-\frac{1}{2}\int d^4x\left[X_i^T(\Box+\mu^2)X_i^T+\chi_n(\Box+\mu^2)\chi_n\right]
\end{equation}
where $\chi_n$ is normalized longitudinal mode, defined as ~\cite{Demozzi2009,Hell2022,Hell_2022,Hell:2021YM}
\begin{equation}
\label{part2-8}
\chi_n=\mu \sqrt{\frac{-\Delta}{-\Delta+\mu^2}}\chi.
\end{equation}
Propagator for longitudinal and transverse modes in position space is given by:
\begin{equation}
\Delta_{\chi_{n}}(x-y) = \int \frac{d^{4} k}{(2 \pi)^{4}} \frac{i}{k^{2} - \mu^2 + i \epsilon} e^{-i k(x-y)}.
\end{equation}
\begin{equation}
\Delta_{i j}^{T}(x-y) = \int \frac{d^{4} k}{(2 \pi)^{4}} \frac{i}{k^{2} - \mu^2 + i \epsilon} \left( \delta_{i j} - \frac{k_{i} k_{j}}{|\vec{k}|^{2}} \right) e^{-i k(x-y)}.
\end{equation}

\noindent In the next sections we use the above splitting of modes.
\section{Unitarity check in massless limit}\label{sec5}

In this section we perform an unitarity analysis of the two-point function in presence of  cubic and quartic self interaction couplings. Which is originally done in ~\cite{Hell2022} for the cubic case.    
\subsection{Cubic self-interaction}
In the hyperbolic formulation, we have seen that adding self-interaction to Proca theory breaks the hyperbolicity of the equation of motion. At the quantum level, unitarity is violated for the self-interacting Proca theory, which is an artifact of perturbation theory, as claimed in~\cite{Hell_2022}. We review the unitarity of the theory at the scale where it loses its hyperbolicity. Let's start with cubic self-interaction  
\begin{equation}
\label{part2-78}
S=\int d^{4} x\left(-\frac{1}{4} F_{\mu \nu} F^{\mu \nu}+\frac{\mu^2}{2} X_{\mu} X^{\mu}-\frac{\lambda}{2} X_{\mu} X_{\nu} X^{\nu, \mu}\right) . 
\end{equation}
The loop correction to the propagator for cubic self-interaction up to two-loop is given by Fig. (\ref{fig:Upto two-loop corrections Upto two-loop corrections}). 
The imaginary part of the one-loop correction is zero~\cite{Hell2022} as shown in Appendix \ref{appendix3.1}. 
The imaginary part of the two-loop diagrams with 3-particle cut contributions is given by~\cite{Hell2022}:
\begin{equation}
\label{part2-91}
\operatorname{Im}\left(\Gamma_{2}\right) = \operatorname{Im}\left(\Gamma_{2 A}\right) + \operatorname{Im}\left(\Gamma_{2 B}\right),
\end{equation}
where
\begin{align*}
\operatorname{Im}\left(\Gamma_{2}\right) &= \int \frac{d^{4} k}{(2 \pi)^{4}} \int \frac{d^{4} l}{(2 \pi)^{4}} \int \frac{d^{4} q}{(2 \pi)^{4}} (2 \pi)^{7} \delta(p-k-l-q) \\
&\times  \delta\left(k^{2}-\mu^2\right)  \delta\left(l^{2}-\mu^2\right) \delta\left(q^{2}-\mu^2\right) \\
&\times \theta\left(q_{0}\right) \theta\left(k_{0}\right) \theta\left(l_{0}\right)\varepsilon_{\mu}(p) \varepsilon_{\nu}(-p) P^{\mu \nu}
\end{align*}
and
\begin{equation}
\label{part2-92}
P^{\mu \nu} \sim -\frac{\lambda^4 l^{\mu} l^{\nu}}{2^5 \mu^{10}} \left[(l q)^4 - 4(l q)^2 (l k)^2 + (l k)^4\right].
\end{equation}

\noindent This represents the most dominant contribution~\cite{Hell2022}.It is immediate that this contribution is singular in mass, signaling a violation of unitarity at two-loops. At first glance, Eq. (\ref{part2-92}) might suggest that the energy scale at which unitarity is violated is \( k_d \sim \lambda^{-\frac{2}{5}} \mu \). While this is nearly correct, \( k_d \) signals the violation of unitarity for the transverse mode\footnote{Polarization vector is independent of the momenta and mass for the transverse mode and does not contribute to scale calculation.} propagator corrections. However, the longitudinal modes, which also show mass-singular behavior, suggest that unitarity is violated at lower scales\footnote{Polarization vector is dependent on the momenta and mass \( k/m \) for the longitudinal mode, which lowers the scale.} \( k_u \sim \lambda^{-\frac{1}{3}} \mu \), in agreement with the findings of ~\cite{Hell2022,HeisenbergZosso2021}.
\subsection{Quartic self-interaction}
After doing a similar exercise for the quartic self-interaction Eq. (\ref{part2-92-0}),   
\begin{equation}\label{part2-92-0}
    S=\int d^4x\left[-\frac{1}{4}F_{\mu\nu}F^{\mu\nu}+\frac{\mu^2}{2}X_{\mu}X^{\mu}+\frac{\lambda}{4}\left(X_{\mu}X^{\mu}\right)^2\right].
\end{equation}
The one-loop diagram for the quartic interaction is a tadpole diagram, which does not contribute to the correction to the propagator. The two-loop diagram, which contributes to the correction to the propagator, has a non-zero imaginary part, as shown in Fig. (\ref{fig:the two-loop quartic self-interaction}).
By taking the imaginary part of two loop correction, which is given in Appendix \ref{appendix3.2}, we get  
\begin{align}\label{imT2Q}
\operatorname{Im}\left(\Gamma_{2 }\right) & =\int \frac{d^{4} p}{(2 \pi)^{4}} \int \frac{d^{4} q}{(2 \pi)^{4}} \int \frac{d^{4} r}{(2 \pi)^{4}}(2 \pi)^{7} \delta(k-p-q-r)\nonumber \\
    & \times \theta\left(p_{0}\right) \delta\left(p^{2}-\mu^2\right)\left(q_{0}\right) \delta\left(q^{2}-\mu^2\right)\theta\left(r_{0}\right) \delta\left(r^{2}-\mu^2\right) \nonumber \\
    & \times \varepsilon_{\mu}(k) \varepsilon_{\nu}(-k) S^{\mu \nu}
\end{align}
where
\begin{align}
\label{part2-92-7}
S^{\mu \nu}=&\frac{\lambda^2}{2 \times3!\times2!\times 4^2}\left(\Theta^\mu_\alpha\Theta^c_\beta+\Theta^c_\alpha\Theta^\mu_\beta+\eta^{\mu c}\Theta^b_\alpha\Theta_{b\beta}\right) \nonumber\\
&\times \left(\Theta^\beta_c \eta^{\nu \alpha}+\Theta^\alpha_c \eta^{\nu \beta}+\Theta^\nu_c \eta^{\alpha\beta}\right) \quad \text{and}, \nonumber \\ 
\Theta_{a\alpha}=&\left[-\eta_{a \alpha}+\frac{p_{a} p_{\alpha}}{\mu^2}\right].
\end{align}
The dominant contribution in Eq. (\ref{part2-92-7}) is of order \( \sim O\left(\frac{\lambda^2 k^6}{\mu^6}\right) \). This contribution is singular in mass, indicating that in the massless limit, the imaginary part of the two-loop correction grows without upper bound, potentially surpassing the unitarity limit ~\cite{Martin:1962rt}. This suggests that unitarity is violated at two loops. Initially, one might conclude from Eq. (\ref{part2-92-7}) and Eq. (\ref{imT2Q}) that this corresponds to the energy scale \( k_d \sim O(\lambda^{-\frac{1}{3}} \, \mathrm{\mu}) \). While this is almost correct, \( k_d \) signals the violation of unitarity, but only for the transverse mode propagator corrections. We have also observed that the polarization vectors for the longitudinal modes are mass-singular. Therefore, for longitudinal modes on the external lines, unitarity is violated at lower scales, specifically \( k_u \sim O(\lambda^{-\frac{1}{4}} \, \mathrm{\mu}) \), which aligns with the strong coupling scale for longitudinal modes given in Eq. (\ref{part2-108}) ~\cite{Hell2022}.\\
\subsection{Unitarity Violation and Perturbation Theory}
It is an interesting fact  
that the observed unitarity violation in the massless limit is an artifact of the perturbation theory. Readers may look into the original reference \cite{Hell2022} for the detailed analysis. This argument involves the Vainshtein mechanism and the decoupling of the longitudinal mode. The Vainshtein mechanism defines a scale, namely the Vainshtein
radius $L_V$, at which the nonlinear terms appearing in the equation of motion become comparable to the linear ones. In the present case, the same also signifies the strong coupling scale, where the perturbation theory breaks down. At this scale the minimal quantum fluctuations of the longitudinal mode  become strong and it decouples from all other degrees of freedoms and remain so for all length scales $L<L_v$. We re-examine the massless limit at energy scales at which the hyperbolicity of the equation of motion is violated.\\

\noindent The minimum quantum fluctuation helps in determining the scale at which regular perturbation theory fails. For $L \sim \frac{1}{k}$ and $\frac{1}{L} \gg \mu$ those are given by:
\begin{equation}
\label{part2-95}
\delta X_{L}^{T} \sim \frac{1}{L}, \quad \delta \chi_{n L} \sim \frac{1}{L} 
\end{equation}
where $\delta X_{L}^{T}$ and $\chi_{n L}$ are the minimal level of quantum fluctuations for the transverse and normalised longitudinal modes. Taking into account the normalised longitudinal mode expressed in terms of the original mode we get,
\begin{equation}
\label{part2-96}
\delta \chi_{L} \sim \frac{1}{L \mu} . 
\end{equation}

\noindent We now find the strong coupling scale of the theory for the longitudinal mode.
Clearly in the case of the normalised longitudinal modes, for a small mass, the first term in Eq. (\ref{part2-107}) of Appendix \ref{appendix4}, is the most dominant one. The scale at which the self-interaction term  from the interacting Lagrangian becomes of the same order as the kinetic term for the longitudinal modes, resulting into their strong coupling scale, is given by 

\begin{equation}\label{part2-108}
    L_v \sim L_{str}\sim\frac{\lambda^{\frac{1}{4}}}{\mu}.
\end{equation}
At this scale, the longitudinal modes enter the strong coupling regime, as $\delta \chi_{L} \sim \frac{1}{\lambda^{\frac{1}{4}}}$ ~\cite{Dvali2007,Hell2022}. Let us now study the same for the transverse mode. The most significant contributor to the transverse modes is the second term. It is of the same order as the kinetic term of the transverse modes at scales
\begin{equation}\label{part2-109}
    L^T_{str}\sim\frac{\lambda^{\frac{1}{3}}}{\mu}, \quad L^T_{str} < L_{str}.
\end{equation}
From Eq. (\ref{part2-108}) and Eq. (\ref{part2-109}) it is clear that the longitudinal modes enter the strong coupling regime first, and the strong coupling scale coincides with the scale  \( k_u \sim O(\lambda^{-\frac{1}{4}} \, \mathrm{\mu}) \) where unitarity seems to violate. Hence, below the strong scale of the longitudinal mode, we cannot trust the perturbation theory for the transverse mode and the corresponding unitarity calculation. To understand the true behavior of the system, we focus on a scale $L$ such that $L^T_{str} < L < L_{str}$. We take this study in the next section.

\section{Decoupling of Longitudinal Modes}\label{sec6}
In this Section we will review the claim decoupling of longitudinal modes beyond the strong coupling regime~\cite{Hell2022}.
Beyond the strong coupling regime, the kinetic term is suppressed by the interaction term, and we need a new canonical normalization that can modify the minimal quantum fluctuations for the longitudinal mode. Once the longitudinal modes enter the strong coupling regime, the most dominant term for the longitudinal modes in the Lagrangian is ~\cite{Hell_2022,Hell2022}

\begin{equation}\label{part2-117}
    \mathcal{L}_{\chi int} = \frac{\lambda}{4} \left( \chi_{,\mu} \chi^{,\mu} \right)^2.
\end{equation}

\noindent This term indicates the new canonical normalization, where the canonically normalized variable is expressed as

\begin{equation}\label{part2-118}
    \bar\chi_{n} \sim \frac{\lambda^{\frac{1}{2}}}{\sqrt{2}} \frac{\chi^2}{L}.
\end{equation}

\noindent Therefore, the dominant term can be rewritten as

\begin{equation}\label{part2-119}
     \mathcal{L}_{\chi int} \sim \frac{1}{2} \bar\chi_{n,\mu} \bar\chi_n^{,\mu}.
\end{equation}

\noindent Using the minimal quantum fluctuations for the normalized mode (\( \delta \bar\chi_n \sim \frac{1}{L} \)), the minimal quantum fluctuations for the original mode is given by

\begin{equation}\label{part2-120}
    \delta_L \chi \sim \frac{1}{\lambda^{\frac{1}{4}}}.
\end{equation}

\noindent Given \( \lambda^{\frac{1}{2}} < 1 \) with the assumption on scales, the condition in Eq.~(\ref{part2-101}) holds automatically.
The minimal quantum fluctuations of the longitudinal mode beyond the strong coupling regime are independent of scale, as shown in~\cite{Hell2022}. 
Hence, the longitudinal modes are decoupled from the theory.
\begin{figure}[H]
            \centering
            \includegraphics[page=12, clip, trim=0cm 10cm 2cm 10cm, width=50mm]{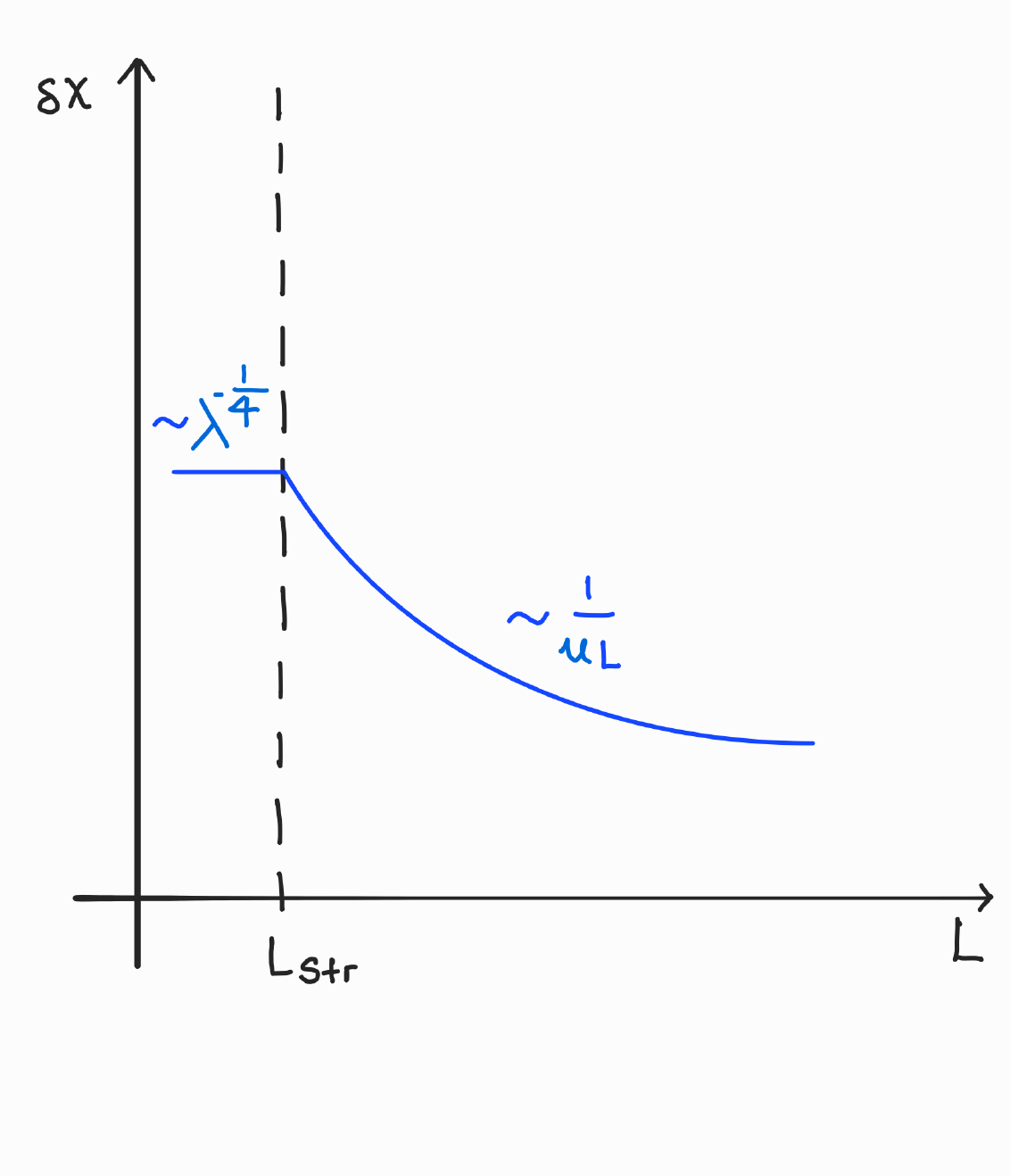}
            \caption{The minimal level of quantum fluctuations of the longitudinal mode in Proca theory with a quartic self-interaction~\cite{Hell2022}.}
            \label{fig:proca_fluctuations}
\end{figure}
\noindent Even though the perturbation theory for longitudinal modes breaks down for scales \( L \leq L_{str} \), it is still applicable for the transverse modes~\cite{Hell_2022}. The most relevant terms for the transverse modes are given by

\begin{equation}\label{part2-121}
    \begin{split}
        & \mathcal{L}_{X^T} \sim -\frac{1}{2} X_i^T (\Box + \mu^2) X_i^T - \lambda \chi_{,\mu} \chi^{,\mu} \chi_{,i} X_i^T \\
        & \quad + \mathcal{O} \left( \lambda \frac{\chi^2}{L^2} \left( X^T \right)^2 \right).
    \end{split}
\end{equation}

\noindent Using Eq. (\ref{part2-120}), the dominant interaction term of the transverse mode for scale $L > L_{str}$ remains subdominant to its kinetic term~\cite{Hell_2022}. Once the longitudinal mode enters the strong coupling regime, the corrections to the transverse modes scale as
\begin{equation}\label{part2-122}
    \lambda \chi_{,\mu} \chi^{,\mu} \chi_{,i} X_i^T \sim \frac{\lambda^{\frac{1}{4}}}{L^4}.
\end{equation}

\begin{equation}\label{part2-123}
    X_i^{T(1)} \sim \frac{\lambda^{\frac{1}{4}}}{L}.
\end{equation}

\begin{figure}[H]
\centering
\includegraphics[page=13, clip, trim=0cm 10cm 1cm 10cm, width=50mm]{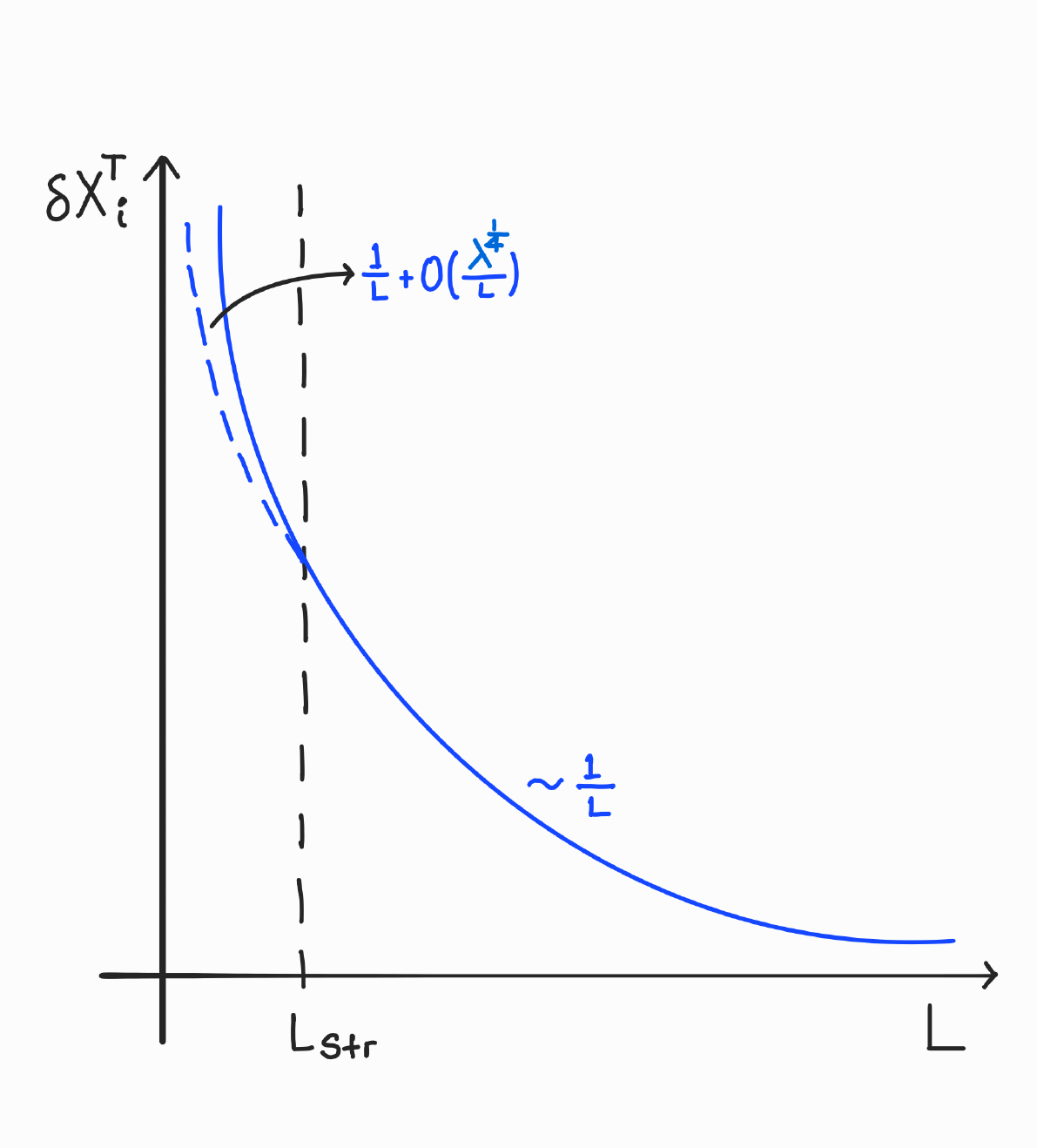}
\caption{The minimal level of quantum fluctuations of the transverse mode in Proca theory with a quartic self-interaction~\cite{Hell2022}.}
 \label{fig:proca_fluctuations1}
\end{figure}

\noindent In self-interacting vector field theories with mass added by hand, divergences in the massless limit are resolved due to the decoupling of the longitudinal mode. Below the strong coupling scale, only the transverse modes remain dynamical. However, as the mass goes to zero, the strong coupling scale moves to infinity, restoring the massless theory in the limit~\cite{Hell2022,Hell_2022}. This behavior, similar to the Vainshtein mechanism in massive gravity, suppresses the longitudinal mode, which becomes non-propagating once it enters the strong coupling regime~\cite{DeFelice2016,Hell2022,Hell_2022}.\\

\noindent For the self-interacting vector field theory without the mass term, the unitarity calculation shows no divergence\footnote{Due to the absence of $0(k^0)$ a term in the propagator of the massless theory, which causes the problem in the massive case.}. If the decoupling persists for the whole region, which is beyond the strong coupling scale length, then we recover the massless theory, and hence the limit is smooth at the quantum level.\\ 

\noindent \textit{Does this decoupling of the longitudinal modes hold for all scales that are beyond the strong coupling scale?} \\

\noindent The claim in ~\cite{Hell2022} regarding the the continuity of the massless limit and the unitarity of the theory is under the assumption that the if the energy scale is much much greater than the mass. However if this assumption is relaxed then as discussed, the breakdown of perturbative analysis cannot be resolved by including non-linear terms (as in the Vainshtein mechanism).

\noindent This motivates us to check the formulation near the scale where the hyperbolicity of the equation is lost, and which is beyond the strong coupling scale.\\

\section{Scale of Hyperbolicity Loss and Constraint Behavior}\label{sec7}
\noindent The claim in \cite{Hell2022} regarding the the continuity of the massless limit and the unitarity of the theory is under the assumption that the the energy scale is much much greater than the mass $(\mu)$. However if this assumption is relaxed, we have already seen that all the pathology discussed in section (\ref{sec2}) arise at the same scale.  In this section we will carefully check the non-covariant formulation near the same scale.\\

\noindent The minimal quantum fluctuations in the covariant formalism can be determined using the equal-time two-point function~\cite{Mukhanov:2007zz}, which is given by 
\begin{align*}
\zeta_{X}(|\vec{x}-\vec{y}|) & \equiv\langle 0| \hat{X}_{\mu}(\vec{x}, t) \hat{X}^{\mu}(\vec{y}, t)|0\rangle .
\end{align*}
The minimal level of quantum fluctuations for the covariant field at scales $\frac{1}{L} \sim \mu$ \footnote{The minimal quantum fluctuation is given by
$\delta X_L \sim \sqrt{\frac{k^3}{\omega}}, \text{ where } \omega = \sqrt{k^2 + {\mu^2}}$ ~\cite{Mukhanov:2007zz}
\text{ then for the case when } $k \sim\frac{1}{L} \sim \mu$ \text{ which is different from the scale used in~\cite{Hell2022}, however the form of minimal} \text{quantum fluctuations remain same.}}
 is given by~\cite{Mukhanov:2007zz}:
\begin{equation}
\label{part2-124}
\delta X_{L} \sim \frac{1}{L} .
\end{equation}
In the perturbative analysis of quartic self-interacting theory, we have seen that the Regular perturbation theory fails, and the scale at which it fails can be calculated via minimal fluctuation
\begin{align}\label{part2-125}
  \mu^2 X_\mu X^\mu &\sim  \lambda(X_\mu X^\mu)^2\nonumber\\
   |X|&\sim \frac{\mu}{\lambda^{\frac{1}{2}}}.
\end{align}
$L_{RPT}$ is the scale at which regular perturbation theory fails 
\begin{align}\label{part2-126}
    L_{RPT}\sim\frac{\lambda^{\frac{1}{2}}}{\mu}.
\end{align}
While doing the perturbative treatment, we showed that the RPT can be restored by removing the secular terms using \textit{Poincaré–Lindstedt method}, but even asymptotic perturbation theory fails when the amplitude of the field becomes 
\begin{align}\label{part2-127}
    \lambda X_\mu X_\nu \sim \mu^2.
\end{align}
The scale at which perturbative treatment can not be defined in the case of quartic self-interaction is 
\begin{align}\label{part2-128}
    L_{Bstr} \sim \frac{\lambda^{\frac{1}{2}}}{\mu}
\end{align}
which coincide with $L_{RPT}$. In the non-covariant formulation, we integrate out the non-propagating part of the vector field using the constraint Eq. (\ref{part2-100})  
\begin{align}\label{part2-129}
    &(-\Delta+\mu^2+\lambda X_iX_i)X_0-\lambda X_0^3=-\Delta\dot{\chi}\nonumber\\
    &(-\Delta+\mu^2-\lambda X_\mu X^\mu)X_0=-\Delta\dot{\chi}\nonumber\\
    &X_0=\frac{-\Delta\dot{\chi}}{(-\Delta+\mu^2-\lambda X_\mu X^\mu)}.
\end{align}
The approximation (\ref{part2-102}) is not well defined at all scales. From Eq. (\ref{part2-129}), it is immediate that the approximation of the constraint (\ref{part2-100}) is violated if $1/L^2 \sim \mu^2 \sim \lambda X^\mu X_\mu$, which happens at the same scale, $L_{Bstr}$.
\begin{figure}[H]
\centering
\includegraphics[page=14, clip, trim=0cm 10cm 1cm 10cm, width=50mm]{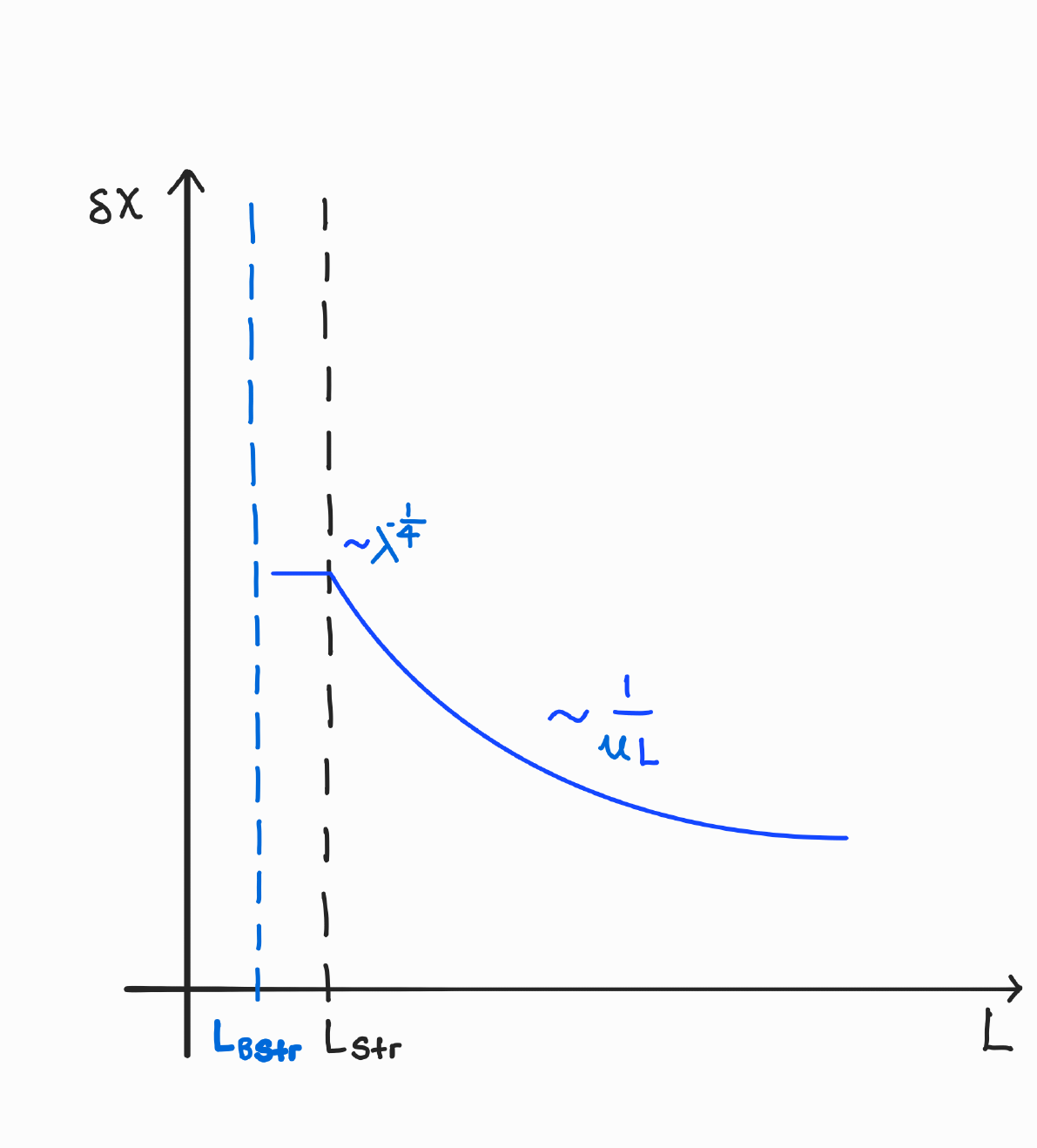}
\caption{The minimal level of quantum fluctuations of the longitudinal mode in Proca theory with a quartic self-interaction.}
 \label{fig:proca_fluctuations2}
\end{figure}

\begin{figure}[H]
    \centering
    \begin{tikzpicture}[scale=1.2, every node/.style={scale=1}]
        \draw[thick] (-5,0) -- (5,0); 

        \draw[thick] (-2,0.2) -- (-2,-0.2);
        \node at (-2,-0.5) {$L_{Bstr} \sim \frac{\lambda^{\frac{1}{2}}}{\mu}$};

        \draw[thick] (2,0.2) -- (2,-0.2);
        \node at (2,-0.5) {$L_{str} \sim \frac{\lambda^{\frac{1}{4}}}{\mu}$};

        \draw[thick] (0,0.2) -- (0,-0.2);
        \node at (0,0.5) {$L^T_{str} \sim \frac{\lambda^{\frac{1}{3}}}{\mu}$};

        \draw[->] (-2,0.4) -- (-3,1);
        \node[align=left] at (-3.8,1.2) {
            \footnotesize Constraint violation Eq. (\ref{part2-100}) and loss of hyperbolicity
        };

        \draw[->] (0,-0.5) -- (0,-1);
        \node[align=left] at (0.5,-1.3) {
            \footnotesize Strong coupling scale for the transverse modes 
            
        };

        \draw[->] (2,0.4) -- (2.8,1);
        \node[align=left] at (3.6,1.3) {
            \footnotesize Longitudinal mode becomes strongly coupled
        };

        \draw[decorate,decoration={brace,amplitude=5pt},yshift=6pt]
            (-2.5,0) -- (-1.5,0) node[midway,yshift=8pt] {\scriptsize \textbf{}};
        \draw[decorate,decoration={brace,amplitude=5pt},yshift=6pt]
            (1.5,0) -- (2.5,0) node[midway,yshift=8pt] {\scriptsize \textbf{}};
        \draw[decorate,decoration={brace,amplitude=5pt,mirror},yshift=-10pt]
            (-0.5,0) -- (0.5,0) node[midway,yshift=-8pt] {\scriptsize \textbf{}};
    \end{tikzpicture}
    \caption{Important scales in the massive self-interacting vector field theory, \( L_{str} > L^T_{str} > L_{Bstr} \)
}
    \label{fig:all-important-scale}
\end{figure}
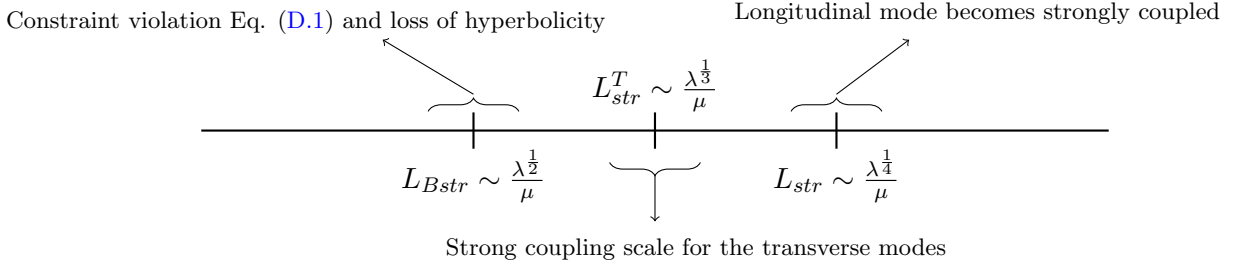

\noindent In figure Fig. \ref{fig:all-important-scale}, we present the three important scales of the theory. As clear from the analysis in this section, \noindent $L_{Bstr}$ is the scale lower than the strong coupling scales for both the longitudinal and transverse modes, where all perturbative methods fail and the non-covariant formulation discussed in the previous section breaks down. This is because of the failure of the approximation (\ref{part2-102}) of the constraint (\ref{part2-100}). If we do not approximate and use the constraint (\ref{part2-100}) to integrate out the non-propagating degree of freedom, the formulation becomes unreliable beyond this scale. The problem of the constraint violation in the non-propagating component of the vector field is also discussed in ~\cite{Constraintvoilation} which coincide with same scale where the system loses it's hyperbolicity.\\

\noindent Now it is clear that the intrinsic pathology we discussed in section (\ref{sec2}) and (\ref{sec3}) is not apparent in this formulation the formulation breaks at the same scale where all those pathology arises.\\

\noindent In conclusion, we find that while the massless theory is well-posed, the massive theory is not well-posed, regardless of how small the mass is. If one uplift the assumption $\frac{1}{L}$ (energy scale) $\gg$ $\mu$ then, the decoupling of the longitudinal mode unreliable below the strong coupling scale and one cannot guarantee the unitarity of the massive self-interacting vector field theory.

\section{Summary and Discussion}\label{sec8}
In this paper we have investigated the classical and quantum pathologies of self-interacting massive vector fields and their massless limits. Since both the cubic and quartic self coupling constants appear with the same mass dimensions, we have incorporated both of them and presented a comparative analysis of various issues in their presence. the primary results are tabulated below.\\

\begin{table}[h!]
\centering 
\begin{tabular}{| p{7cm} | p{7cm}  |}
\hline
\textbf{\,\,\,\,\,\,\,\,\ Quartic self-interaction} & \textbf{\,\,\,\,\,\,\,\,\ Cubic self-interaction} \\
\hline
1. Hyperbolicity of EOM is lost~\cite{PhysRevLett.129.151103}. & 
1. Hyperbolicity of EOM is lost, which is numerically shown in Ref. ~\cite{PhysRevD.108.044022}. \\
\hline
2. Without a mass term, the effective metric is non-degenerate and the theory is well-posed. & 
2. Determinant of the effective metric is unknown\cite{PhysRevD.108.044022}, even in the massless limit — hyperbolicity status is unclear. \\
\hline
3. Loop corrections diverge in the massless limit, attributed to perturbative artifacts in~\cite{Hell2022}. However, the non-covariant formulation breaks at the same scale as hyperbolicity loss, hence decoupling of longitudinal mode at $L_{\text{Bstr}}$ is uncertain. Without the mass term, hyperbolicity is preserved, ensuring unitarity. & 
3. Without a mass term, hyperbolicity is not guaranteed, hence longitudinal mode decoupling is uncertain, unlike the quartic case. 
\\
\hline
\end{tabular}
\caption{Comparison of quartic and cubic self-interactions in terms of hyperbolicity and longitudinal mode behavior at different scales and limits.}
\label{tab:quartic_vs_cubic}
\end{table}

\noindent At the classical level, massive self-interacting vector fields show pathological behavior, namely the singularity formation in 1+1 dimensions and loss of hyperbolicity in dimensions $ D \geq 2 $. We take the massless limit of the quartic self-interacting vector field theory in hyperbolic formulation, which shows that the massless limit is not smooth. Moreover, due to self-interaction, the theory becomes quasilinear, and the perturbative treatment of the theory cannot be defined. Both the perturbative techniques, i.e., regular and asymptotic perturbation theories, fail in this case.\\

\noindent As is well known, quantum corrections can avoid the classical singularity (kink) by smoothing the kink. Hence it is necessary to check whether the pathological behavior at the classical level is carried to the quantum level or not. The conventional approach, in which one computes the imaginary part of the correction to the propagator for a self-interacting vector field, diverges at two-loop levels~\cite{Hell2022}, which indicates a violation of unitarity at two-loop levels. This is also an artifact of the the perturbation theory, which is due to the longitudinal mode of the massive vector fields~\cite{Hell2022}. The claim in~\cite{Hell2022} regarding the continuity of the massless limit and the unitarity of the theory is under the assumption that the the energy scale is much much greater than the mass $(\mu)$. However we wanted to understand the other regime, where they are of the same scale. This is promoted from a renormalization flow perspective, where both regimes (UV and IR) are important. We find that if this assumption of~\cite{Hell2022} is relaxed then as discussed, the breakdown of the perturbative analysis cannot be resolved by including non-linear terms (as in the Vainshtein mechanism) and the non covariant formulation uses a constraint equation in the non-propagating degree of freedom and the propagating degree of freedom to further express the action of the theory completely in terms of the propagating degree of freedom. This feature is violated at the scale $L_{Bstr}$ beyond the strong coupling regime of all the degrees of freedoms of the theory. Since the approximation is violated at the scale $L_{Bstr}$, one cannot use the constraint to express the action in terms of the propagating degree of freedom. $L_{Bstr}$ is the same scale where the hyperbolicity is lost and asymptotic perturbation theory fails. Hence, incorporating nonlinear terms does not help in removing the divergence. To conclude, we find that while the massless theory is well-posed, the massive theory is not well-posed, regardless of how small the mass is. If one uplift the assumption $\frac{1}{L}$ (energy scale) $\gg$ $\mu$ then, the theory is not well-behaved beyond the scale $L_{Bstr}$ for any non zero mass. Moreover in the massless limit one cannot guarantee the unitarity of the massive self-interacting vector field theory. Hence the unitarity of the theory is scale-dependent. When the energy scale becomes comparable to the mass, pathologies can arise, and unitarity is no longer guaranteed.\\
\\
\noindent Let us end the paper with some interesting open questions. We have seen the appearance of different length scales in the NPL theory gives us important physical characteristics. It would be interesting to understand the same via a renormalization group flow study. We shall report on the same in a future work. In this paper, we consider the mass term added by hand to self-interacting vector fields. It is an interesting open problem to find the fate of hyperbolicity and unitarity of the system, if the mass is generated by a symmetry-breaking mechanism. On a different note because of the different behavior of massive abelian self-interacting vector fields and massive non-abelian vector fields (massive Yang-Mills) at the classical level, the unitarity of massive Yang-Mills can not be studied by taking self-interacting abelian vector fields as a toy model, as is done in~\cite{Hell2022}.  There is no violation of hyperbolicity in the equations of motion for massive Yang-Mills theory and consequently, there is no loss of unitarity in this case as discussed in Appendix \ref{appendix5}~\cite{Hell:2021YM}. Does the addition of additional self-interaction terms in massive Yang-Mills theory inevitably lead to a loss of hyperbolicity and, consequently to a violation of unitarity? Under what conditions, if any, can hyperbolicity and unitarity be preserved in such theories?. Exploration of this question is an important part of ongoing research~\cite{jose2025internalsymmetryrescuewellposed} and future research. 

\acknowledgments
We thank Suvankar Dutta, Karan Farnandes, Akshay Kumar, Arpita Mitra and Arnab Rudra for useful discussions on the project. We also thank Dr. Anamaria Hell and the reviewer for their useful comments on the first version of the manuscript. NB would like to acknowledge the
hospitality of ICTP during the final stage of the work. The work of NB is supported by SERB
POWER fellowship SPG/2022/000370 and we thank people of India for their generous support to
the advancement of basic sciences.\\
\appendix
\section{Effective metric for self-interacting vector field without mass}\label{appendix1}
\noindent Equation of motion for Eq. (\ref{action without mass term}) is $$\nabla_\mu F^{\mu \nu}=-\lambda X^2 X^\nu.$$
Recall the generalized Lorenz constraint: $\nabla_\mu \nabla_\nu F^{\mu \nu}=0$
$$
\begin{aligned}
&\Rightarrow \quad \nabla_\mu\left[\lambda X^2 X^\mu\right]=0 \\
& \Rightarrow \quad \nabla_\mu X^\mu=-\frac{X^\mu \nabla_\mu\left(\lambda X^2\right)}{\lambda X^2}.
\end{aligned}
$$
Let $z_0'=-\lambda X^2$, using the Lorenz constraint, the EOM becomes 
\begin{align}
 \Rightarrow &\nabla^\mu F_{\mu \nu}-z_0' X_\nu=0 \\
 \Rightarrow &\nabla^\mu \nabla_\mu X_\nu-\nabla_\nu \nabla_\mu X^\mu-R_{\mu \nu} X^\mu-z_0' X_\nu=0 \\
 \Rightarrow &\nabla^\mu \nabla_\mu X_\nu+\nabla_\nu\left(z_0'^{-1} X^\mu \nabla_\mu z_0'\right)-R_{\mu \nu} X^\mu-z_0' X_\nu=0 \\
 \Rightarrow &(g_{\mu\rho} - 2\lambda z_0'^{-1} X_\mu X_\rho) \nabla^\mu \nabla^\rho X_\nu - 2 \lambda z_0'^{-1}X^\mu X^\rho \nabla_\mu F_{\nu\rho} \nonumber \\
 &- R_{\mu \nu} X^\mu - z_0' X_\nu + \cdots = 0
\end{align}
where ellipses represent the product of single derivative terms.
\begin{align}\label{eq:eeomm0}
    \text{Let, }\bar{g}_{\mu\rho} = g_{\mu\rho} - 2\lambda z_0'^{-1} X_\mu X_\rho \text{ and } \Theta_{\nu}= X^\mu X^\rho \nabla_\mu F_{\nu\rho}\\ 
    \label{eq:eeomm02}
    \bar{g}_{\mu\rho} \nabla^\mu \nabla^\rho X_\nu - 2\lambda z_0'^{-1} \Theta_\nu  - R_{\mu \nu} X^\mu - z_0' X_\nu + \cdots = 0 
\end{align}
\section{Local perturbative analysis}\label{appendix2}
We perform a local analysis in flat spacetime for simplicity, but before that, we should check whether local analysis does not violate any global constraint and has a solution for the Equation of motion.
Let's take a global time-like vector field $X^{0 \mu}=\left(\tilde{X}^0, 0\right)$, which is the solution of Proca theory.
Using the local Lorenz constraint, the local equation of motion becomes
$$
\begin{aligned}
O\left(\lambda^0\right)&: \square \tilde{X}^0+\mu^2 \tilde{X}^0=0 \Rightarrow \partial_x^2 \tilde{X}^0-\mu^2 \tilde{X}^0=0, \\
O\left(\lambda^{1}\right)&: \square \delta X_v+\mu^2 \delta X_v+J_v=0 \\
&\Rightarrow\left\{\begin{array}{l}
\square \delta X_0+\mu^2 \delta X_0+\tilde{X^0}^{3}=0 \\
\square \delta X_1+\mu^2 \delta X_1=0.
\end{array}\right.
\end{aligned}
$$
For a global space like vector field $X^{0\mu}$ which is a solution of Proca theory, then locally $X^{0 \mu}$ is $\left(0, \tilde{X}^1\right)$. Using the local Lorenz constraint, the local equation of motion becomes
$$
\begin{aligned}
 O\left(\lambda^0\right)&:\partial_t^2 \tilde{X}^{1}+\mu^2 \tilde{X}^1=0, \\
 O\left(\lambda^{1}\right)&: \square \delta X_v+\mu^2 \delta X_v+J_v =0 \\
&\Rightarrow\left\{\begin{array}{l}
\square \delta X_0+\mu^2 \delta X_0=0 \\
\square \delta X_1+\mu^2 \delta X_1+\tilde{X^1}^{3}=0.
\end{array}\right.
\end{aligned}
$$
\subsubsection{Poincare-Lindstedt method}
We use the Poincaré-Lindstedt method to remove the secular term from the solution. Let's define a stretched time variable $\tau$  as
$$
\begin{aligned}
\tau & =\omega(\lambda) t \\
\Rightarrow \frac{d}{d t} & =\omega(\lambda) \frac{d}{d \tau}.
\end{aligned}
$$
Asymptotic expansion of the form
$$
\begin{aligned}
& X^\mu=X^{0\mu}+\lambda \delta X^\mu+O\left(\lambda^2\right), \\
& \omega(\lambda)=1+\lambda \omega_1+ O\left(\lambda^2\right).
\end{aligned}
$$
Solution of $O\left(\lambda^{0}\right): \tilde{X}^{1}=A e^{i \mu t}+B e^{-i \mu t}$.\\
Solution of $O\left(\lambda^{1}\right)$: $\delta X_1=\delta X^0_ {\text{homo}} +\delta X^1_{\text{particular}}$
\begin{align}
\delta X_1 &=A^{\prime} e^{i(\omega t-k x)}+B^{\prime} e^{-i(k x+\omega t)} \nonumber\\
&-\frac{(3 A^2 B +2\mu^2 A \omega_1)}{2 i \mu} t e^{i \mu t}+\frac{A^3}{8 \mu^2} e^{3 i \mu t} \nonumber\\
&+\frac{(3 B^2 A +2\mu^2 B \omega_1)}{2 i \mu} t e^{-i \mu t}+\frac{B^3}{8 \mu^2} e^{-3 i \mu t}.
\end{align}
The secular term :$-\frac{3 A^2 B +2\mu^2 A \omega_1}{2 i \mu} t e^{i \mu t}$ and $\frac{3 B^2 A +2\mu^2 B \omega_1}{2 i \mu}$ $ t e^{-i \mu t}$.\\
To remove the secular term  $\frac{3 A B}{2 \mu^2}+ \omega_1$ should be zero and for that $\omega_1=-\frac{3 A B}{2 \mu^2}$.
\section{Loop correction for self-interacting vector field theories}\label{appendix3}
\subsubsection{Cubic self-interaction}\label{appendix3.1}
\begin{figure}[H]
\begin{center}
\includegraphics[page=2, clip, trim=2cm 28.5cm 2cm 3cm, width=40mm]{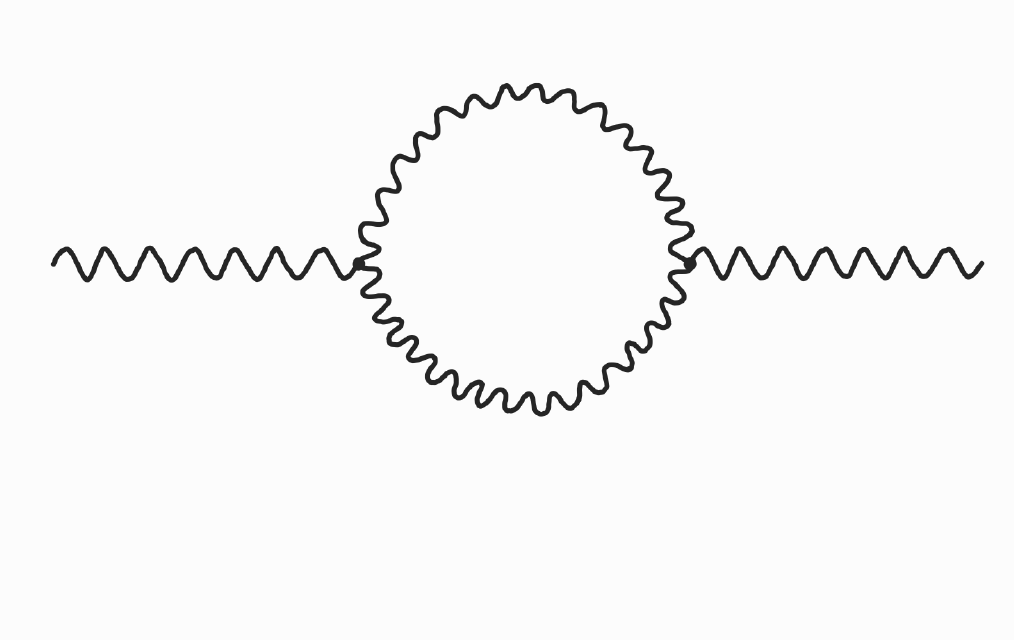}
\includegraphics[page=2, clip, trim=2cm 18cm 2cm 14cm, width=40mm]{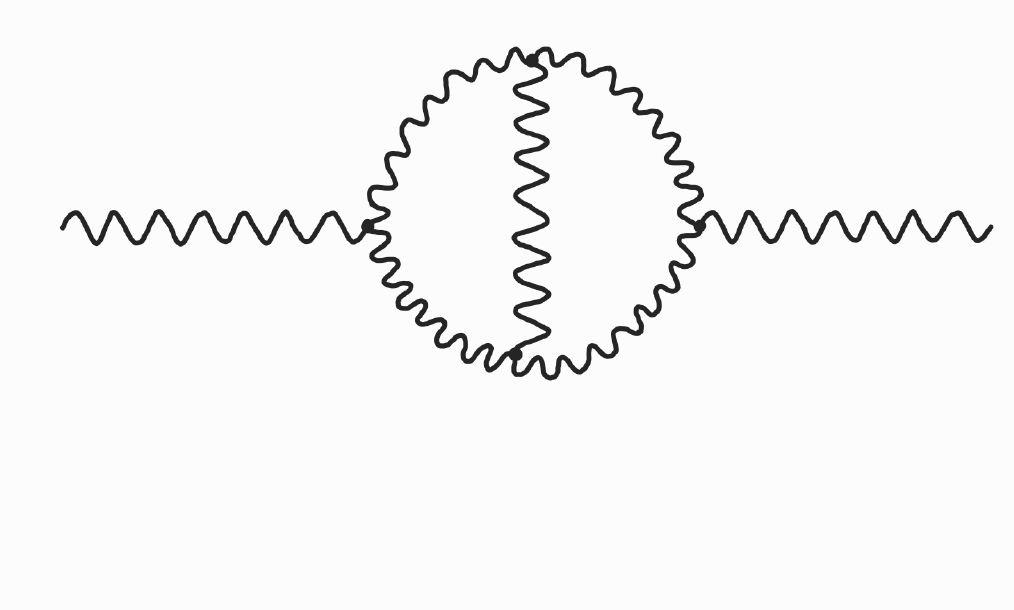}
\includegraphics[page=2, clip, trim=2cm 6.5cm 2cm 24cm, width=40mm]{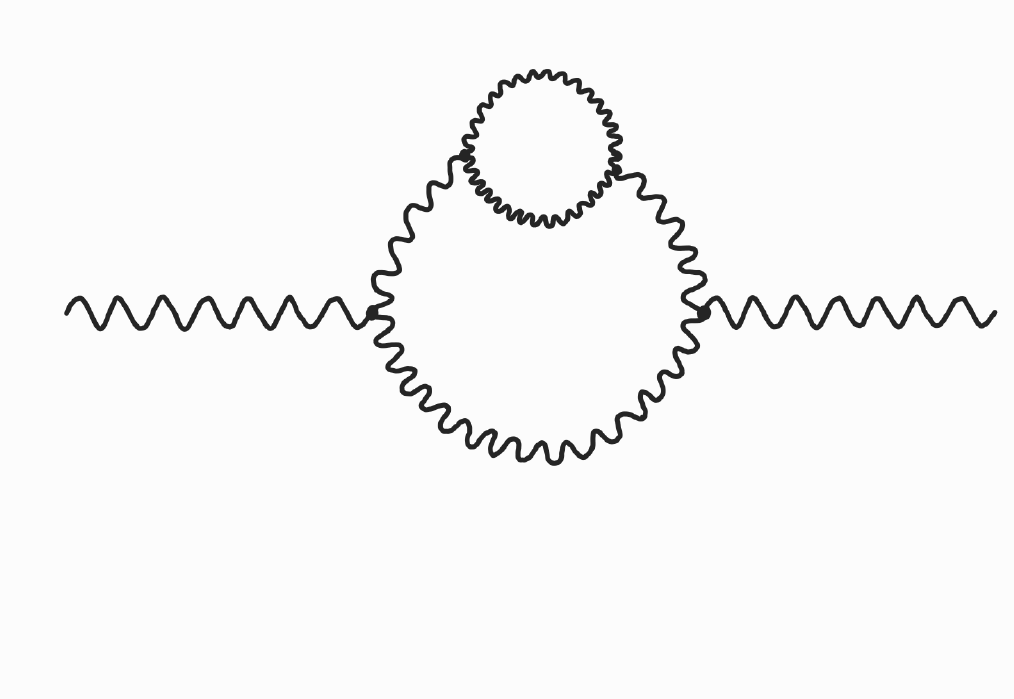}
\end{center}
\caption{Upto two-loop corrections of the propagator~\cite{Hell2022}.} 
\label{fig:Upto two-loop corrections Upto two-loop corrections} 
\end{figure}

The loop correction to the propagator for cubic self-interaction up to two-loop is given by Fig. (\ref{fig:Upto two-loop corrections Upto two-loop corrections}) 
\noindent The imaginary part of the one-loop correction is given by ~\cite{Hell2022}:
\begin{align*}
\operatorname{Im}\left(\Gamma_{1}\right) =& \varepsilon_{\mu}(p) \varepsilon_{\nu}(-p) \int \frac{d^{4} k}{(2 \pi)^{3}} \int \frac{d^{4} q}{(2 \pi)^{3}} (2 \pi)^{4} \delta(q+k-p) \\
& \times \theta\left(k_{0}\right) \theta\left(q_{0}\right) \delta\left(k^{2}-\mu^2\right) \delta\left(q^{2}-\mu^2\right) L^{\mu \nu}
\end{align*}
where
\begin{align}
\label{part2-86}
&L^{\mu \nu} = \frac{\lambda^{2}}{16} V^{\mu \alpha \beta}(-p, k, q)\left(-\eta_{\alpha \gamma}+\frac{k_{\alpha} k_{\gamma}}{\mu^2}\right)\left(-\eta_{\beta \delta}+\frac{q_{\beta} q_{\delta}}{\mu^2}\right) \nonumber\\
&V^{\mu \nu \gamma}(k, p, q) = i\left(k^{\mu} \eta^{\nu \gamma} + p^{\nu} \eta^{\mu \gamma} + q^{\gamma} \eta^{\mu \nu}\right).
\end{align}
Because of the delta functions, the internal momenta are on-shell, i.e. \( k^2 = q^2 = \mu^2 \). Then, noting that
\begin{equation}
\label{part2-87}
k^{\gamma}\left(-\eta_{\alpha \gamma} + \frac{k_{\alpha} k_{\gamma}}{\mu^2}\right) = 0 \quad \text{for} \quad k^2 = \mu^2,
\end{equation}
it follows that \( L_{\mu \nu} = 0 \) ~\cite{Hell2022}. Therefore, at one loop, there is no discontinuity. 
\begin{figure}[H]
\begin{center}
\includegraphics[page=4, clip, trim=2cm 9.5cm 2cm 24cm, width=50mm]{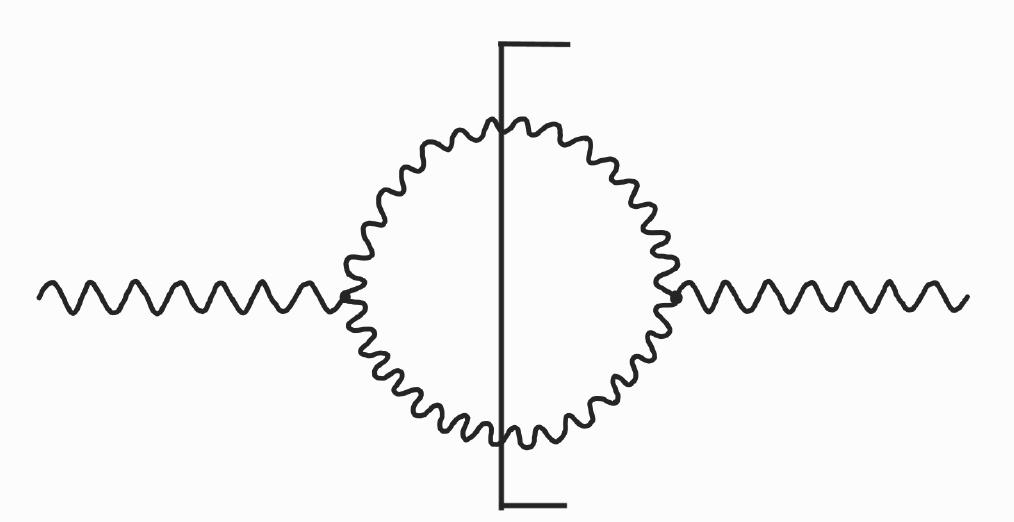}
\end{center}
\caption{The cut of the one-loop corrections to the propagator for the cubic self-interaction~\cite{Hell2022}.}
\label{fig:cropped_pdfs}
\end{figure}
\subsubsection{Quartic self-interaction}\label{appendix3.2}
\begin{figure}[H]
\begin{center}
\includegraphics[page=8, clip, trim=2cm 10cm 3cm 21cm, width=50mm]{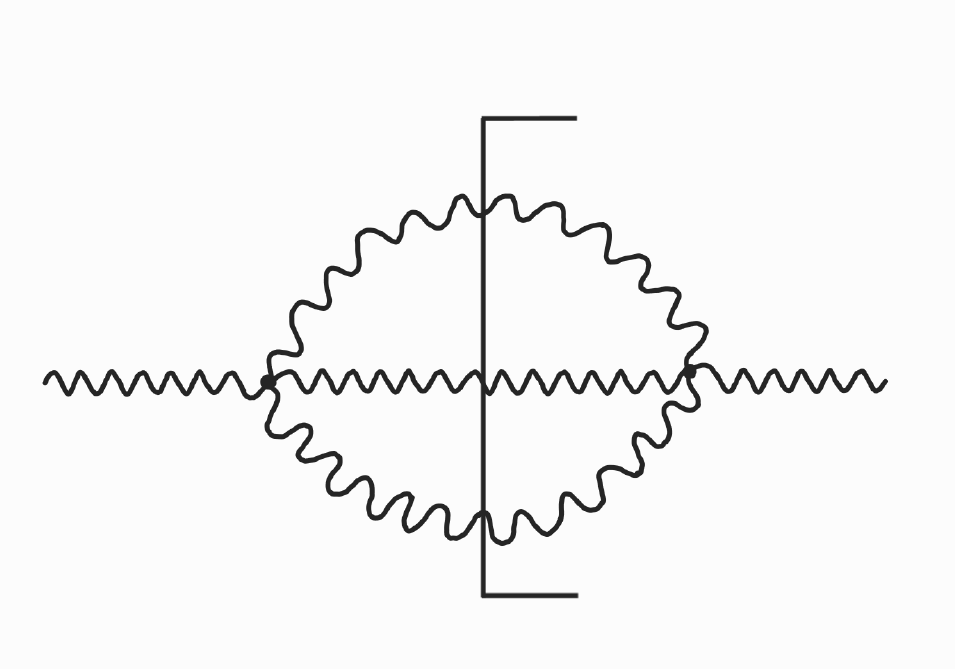}
\end{center}
\caption{The Cuts of the two-loop corrections to the propagator in case of quartic self-interaction.} 
\label{fig:the two-loop quartic self-interaction} 
\end{figure}

The one-loop diagram for the quartic interaction is a tadpole diagram, which does not contribute to the correction to the propagator. The two-loop diagram, which contributes to the correction to the propagator, has a non-zero imaginary part; is Fig. (\ref{fig:the two-loop quartic self-interaction})
\begin{figure}[H]
\begin{center}
\includegraphics[page=5, clip, trim=2cm 18cm 2cm 15cm, width=60mm]{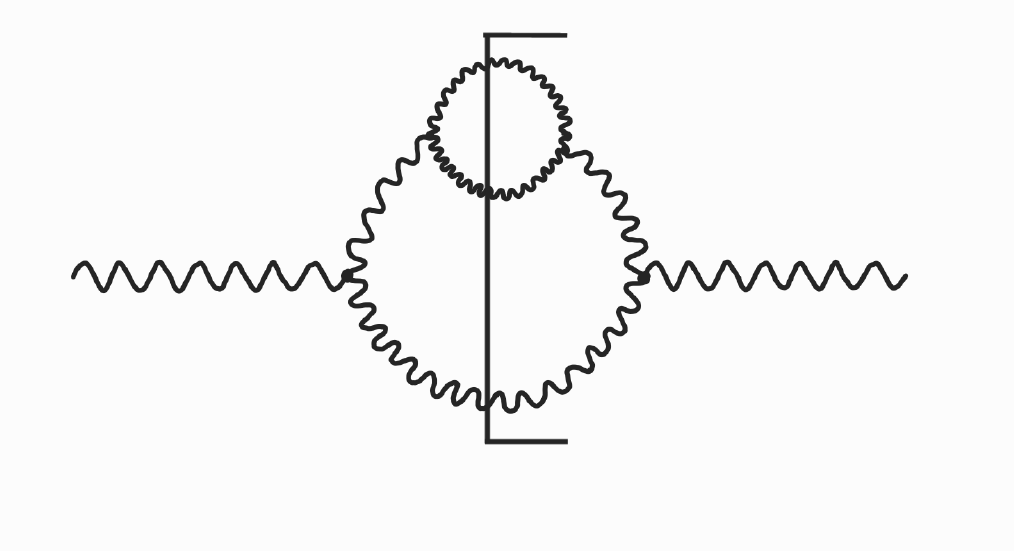}\\
\includegraphics[page=5, clip, trim=2cm 10cm 2cm 28cm, width=80mm]{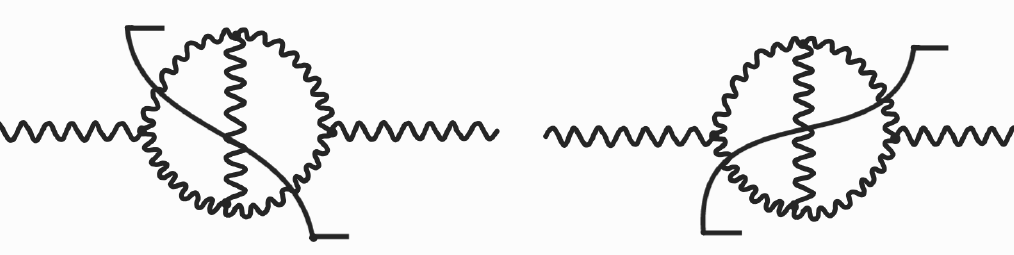}
\end{center}
\caption{The cuts of the two-loop corrections to the propagator for the cubic self-interaction~\cite{Hell2022}.}
\label{fig:cropped_pdfs}
\end{figure}
\noindent The two-loop correction from the sunset diagram is given by
\begin{align}\label{part2-92-3}
    i\Gamma_{2 }=& \frac{1}{3!2!}\int \frac{d^{4} p}{(2 \pi)^{4}} \int \frac{d^{4} q}{(2 \pi)^{4}} \int \frac{d^{4} r}{(2 \pi)^{4}}(2 \pi)^{4} \delta(k-p-q-r)\nonumber \\
    & \times V^{\mu a b c} \Delta_{a \alpha}(p) \Delta_{b \beta}(q) \Delta_{c \gamma}(r) V^{\nu \alpha \beta \gamma} \varepsilon_{\mu}(k) \varepsilon_{\nu}(-k)
\end{align}
where \begin{align*}
    V^{\mu \nu \rho \sigma}=& \lambda \left( \eta^{\mu \nu}\eta^{\rho \sigma}+\eta^{\mu \rho}\eta^{\nu \sigma}+\eta^{\mu \sigma}\eta^{\nu \rho}\right) \text{  and,} \\
    \Delta_{\mu \nu}(k) =& \left(-\eta_{\mu \nu} + \frac{k_{\mu} k_{\nu}}{\mu^2}\right) \frac{i}{k^{2} - \mu^2}.
\end{align*} 
\section{Strong coupling regime for quartic self-interaction}\label{appendix4}
For the case of quartic self-interaction, varying the action (\ref{part2-92-0}) with respect to $X_0$, we find a constraint satisfied by $X_0$ ~\cite{Hell2022}
\begin{equation}\label{part2-100}
    (-\Delta+\mu^2+\lambda X_iX_i)X_0-\lambda X_0^3=-\Delta\dot{\chi},
\end{equation}
where $X_i$ is given by Eq. (\ref{part2-3}). Assuming
\begin{equation}\label{part2-101}
    \lambda \chi^2<1,
\end{equation}
which should be verified a posteriori, we can resolve this constraint for scales $k^2\sim\frac{1}{L^2}\gg \mu^2$ as
\begin{equation}\label{part2-102}
\begin{split}
    X_0&=\frac{-\Delta}{-\Delta+\mu^2}\dot{\chi}+\frac{\lambda}{-\Delta+\mu^2}\\&\times\left[\left(\dot{\chi}^2-\chi_{,i}\chi_{,i}-2\chi_{,i}X_i^T-X_i^TX_i^T\right)\dot{\chi}\right]\\
    &+\mathcal{O}\left(\lambda(\mu L)^2\frac{\chi^3}{L},\lambda^2\frac{\chi^5}{L}\right).
\end{split}
\end{equation}
Here $\frac{1}{L}$ represents a derivative acting on $\chi$. Substituting this expression back into the action, we obtain the Lagrangian density
\begin{equation}\label{part2-103}
    \mathcal{L}=\mathcal{L}_0+\mathcal{L}_{int} .
\end{equation}
In $\mathcal{L}_{int}$, we have retained only the most relevant terms ~\cite{Hell2022}. Using the normalized variables $\chi_n$ defined in Eq. (\ref{part2-8}), for scales $\frac{1}{L^2}\gg \mu^2$, we obtain
\begin{equation}
    \begin{split}
        \mathcal{L}_0&=-\frac{1}{2}\chi_n(\Box+\mu^2)\chi_n-\frac{1}{2}X_i^T(\Box+\mu^2)X_i^T\quad\text{and,}\\
        \mathcal{L}_{int}&\sim\frac{\lambda}{4\mu^4}\left(\chi_{n,\mu}\chi_n^{,\mu}\right)^2-\frac{\lambda}{\mu^3}\chi_{n,\mu}\chi_n^{,\mu}\chi_{n,i}X_i^T\\ 
        &-\frac{\lambda}{2\mu^2}\chi_{n,\mu}\chi_n^{,\mu}X_i^TX_i^T+\frac{\lambda}{\mu^2}\left(\chi_{n,i}X_i^T\right)^2.
    \end{split}\label{part2-106}
\end{equation}
Using the minimal level of quantum fluctuations, the Vainshtein radius is determined by the scale where the most dominant term is of the same order as the kinetic term. Assuming the derivatives as $\partial_{\mu}\sim\frac{1}{L}$ and the normalised longitudinal and transverse modes as $\chi_n\sim\frac{1}{L}$ and $X_i^T\sim\frac{1}{L}$, the order of interaction terms is given as follows:
\begin{equation}\label{part2-107}
\begin{split}
     \frac{\lambda}{4\mu^4}\left(\chi_{n,\mu}\chi_n^{,\mu}\right)^2&\sim\frac{\lambda}{(\mu L)^4L^4},\\ \frac{\lambda}{\mu^3}\chi_{n,\mu}\chi_n^{,\mu}\chi_{n,i}X_i^T&\sim\frac{\lambda}{(\mu L)^3L^4}\quad\text{and,}\\
     \frac{\lambda}{2\mu^2}\chi_{n,\mu}\chi_n^{,\mu}X_i^TX_i^T&\sim\frac{\lambda}{\mu^2}\left(\chi_{n,i}X_i^T\right)^2\sim\frac{\lambda}{(\mu L)^2L^4}.
\end{split}
\end{equation}

\section{Massive Yang-Mills}\label{appendix5}
The action for massive Yang-Mills is given by
\begin{align}
    S =\int \sqrt{-g}\, d^4x[- \frac{1}{4}F_{a\mu\nu}F^{a\mu\nu} + \frac{1}{2}\mu^{2} B^{\mu}_{a}B_{\mu}^{a}
   ],\label{eqn:action}
\end{align}
Here, $B_\mu^a$ is a vector field in the Lie algebra of SU(2), $F_{a\mu\nu} = \nabla_\mu B_{a\nu} - \nabla_\nu B_{a\mu} + g \epsilon_{abc} B^b_\mu B^c_\nu$ is the field strength tensor, $g$ is the gauge coupling, $\epsilon_{abc}$ the SU(2) structure constants, and $\mu$ the common mass of the vector fields. It is worth noting that the Yang-Mills theory is inherently self-interacting. Specifically, it exhibits two types of self-interactions: quartic self-interactions of the form $ g^2 \epsilon_{abc} \epsilon^{ade} B_{\mu}^{b} B_{\nu}^{c} B^{\mu}_{d} B^{\nu}_{e}$,
and derivative self-interactions of the form $g \epsilon^{abc} (\nabla_\mu B_{a\nu}) B^\mu_b B^\nu_c$.\\

\noindent Introducing the gauge covariant derivative $D_{\alpha} F_{a\mu }^{\alpha }  \equiv \nabla_{\alpha} F_{a\mu }^{\alpha } + g B^{b\alpha } \epsilon_{abc} F^{c}_{\mu \alpha}$, the equation of motion can be written as
\begin{equation}
       D_{\alpha }F_{a\mu }^{\alpha } = \mu^2 B_{a\mu } \,.\label{eqn:eqmotion}
\end{equation}
Due to the symmetry of the field strength tensor, the condition $D^{\mu}D_{\alpha }F_{a\mu }^{\alpha }=0$ holds. The action \eqref{eqn:action} is invariant under internal global SU(2) transformations, which are given by
\begin{equation}
    B_{a\mu} \mapsto B_{a\mu} + \delta B_{a\mu} = B_{a\mu} - i \epsilon_{abc} \lambda^b A^{c}{}_\mu, \label{eqn:globalsu2}
\end{equation}
where $\lambda^{b}$ are constants and the conserved currents are
\begin{equation}\label{sourceMYM}
    J^{a\nu} = F^{b\nu\mu} \epsilon^{a}{}_{bc} B^c{}_{\mu}.
\end{equation}
Applying the gauge covariant derivative to the equation of motion leads to the Lorenz condition
\begin{equation}
    \nabla^{\mu}\left(z_{ab} B^{b}_{\mu} \right)=0\,,\label{eqn:lorentzgen}
\end{equation}
where $z_{ab}\equiv {\mu}^{2}g_{ab}$. In general, the equation of motion Eq.~\eqref{eqn:eqmotion} can be written in the reduced form
\begin{equation}
     D_{\alpha }F_{a\mu }^{\alpha}=z_{ab} B_{\mu}^{b}\,.
\end{equation}
For massive Yang-Mills, the principal part of the equation of motion is field independent unlike the abelian self-interacting (quartic or cubic self-interaction) vector field this is due to the internal symmetry which helps in getting rid of the second term of gauge covariant derivative, which contribute in  principal part by modifying the Lorenz condition. In conclusion mass added by hand is not problematic in case of massive Yang-Mills~\cite{jose2025internalsymmetryrescuewellposed} and massive self-interacting abelian vector field suffers from the pathological behavior and resolve in absence of the mass term.
Unlike the self-interacting abelian vector field, where the approximation the constraint Eq. (\ref{part2-100}) is not well-defined at the scale where hyperbolicity of EOM loses and the signature of kinetic term for the longitudinal mode changes beyond the scale. In case of massive Yang-Mills, whose equation of motion is three copies of Proca field with a conserved source, given by Eq. (\ref{sourceMYM}). The constraint equation and the action in terms of propagating degree of freedom is given by~\cite{Hell2022}  
\begin{align}\label{eq::constraint}
    (-\Delta+\mu^2)B_0^a=&-\dot{B}_{i,i}^a-g\epsilon^{abc}\dot{B}_i^bB_i^c-g\epsilon^{abc}B_i^bB_{0,i}^c\nonumber\\
    &-g\epsilon^{abc}\partial_i(B_i^bB_0^c)-g^2\epsilon^{fbc}\epsilon^{fad}B_0^bB_i^cB_i^d,
\end{align}
the non-propagating component of the field up to $\mathcal{O}\left(g^2\right)$ can be approximated as
\begin{equation}\label{eq::constraintsol}
\begin{split}
    B_0^a=&D\left[\dot{\chi}^a\right]
    -\frac{g\varepsilon^{abc}}{-\Delta+\mu^2}\left[\dot{B}_i^bB_i^c+(\Delta\chi^b+2B_i^b\partial_i)D\left[\dot{\chi}^c\right]\right]\\
    &+\frac{g^2\varepsilon^{fab}\varepsilon^{fcd}}{-\Delta+\mu^2}(\Delta\chi^b+2B_i^b\partial_i)\frac{1}{-\Delta+\mu^2} \times \\
    &\left[\dot{B}_j^cB_j^d+(\Delta\chi^c+2B_j^c\partial_j)D\left[\dot{\chi}^d\right]\right]\\
    &+\frac{g^2\varepsilon^{fab}\varepsilon^{fcd}}{-\Delta+\mu^2}B_i^bB_i^cD\left[\dot{\chi}^d\right].
\end{split}
\end{equation}
Here, $\chi^a$ is the longitudinal component of $B_i^a$, that we decompose according to 
\begin{equation}\label{eq::decomposition}
    B_i^a=B_i^{Ta}+\chi_{,i},\qquad\text{where}\qquad B_{i,i}^{Ta}=0
\end{equation}
while the operator $D$ is given by 
\begin{equation}
    D[\chi]=\frac{-\Delta}{-\Delta+\mu^2}\chi.
\end{equation}
Integrating out the non-propagating, we obtain the Lagrangian density 
\begin{equation}\label{eq::Llin}
    \mathcal{L}=\mathcal{L}_0+\mathcal{L}_{int},\qquad\text{where}
\end{equation}
\begin{equation*}
    \begin{split}
        \mathcal{L}_0=&-\frac{1}{2}\chi^a(\Box+\mu^2)\frac{\mu^2(-\Delta)}{-\Delta+\mu^2}\chi^a-\frac{1}{2}B_i^{Ta}(\Box+\mu^2)B_i^{Ta}\\        
        \mathcal{L}_{int}=&g\varepsilon^{abc}\left[\frac{1}{2}\chi^b\chi^c_{,i}\Box B_i^{Ta}-\dot{\chi}^a\chi^{c}_{,i}\frac{\mu^2}{\Delta}\left(\dot{\chi}^b_{,i}\right)+\chi^bB_i^{Tc}\Box B_i^{Ta}\right]\\
        &+\frac{g^2}{2}\varepsilon^{fab}\varepsilon^{fcd}\dot{\chi}^a\dot{\chi}^c\chi^b_{,i}\chi^{d}_{,i}\\
        &+\frac{g^2}{2}\varepsilon^{fab}\varepsilon^{fcd} \left(\dot{\chi}^a_{,i}\chi^b_{,i}+\dot{\chi}^a\Delta\chi^b\right)\frac{1}{\Delta}\left(\dot{\chi}^c_{,j}\chi^d_{,j}+\dot{\chi}^c\Delta\chi^d\right)\\
        &-\frac{g^2}{4}\varepsilon^{fab}\varepsilon^{fcd}\chi^a_{,i}\chi^c_{,i}\chi^b_{,j}\chi^d_{,j}\\
        &+\mathcal{O}\left(\frac{gB^{T3}}{L},\frac{g \mu^2 B^T\chi^2}{L^3}, \frac{g^2B^T\chi^3}{L^3}\right).
    \end{split}
\end{equation*}

\bibliographystyle{JHEP}
\bibliography{refs}


\end{document}